\documentclass{nature_incl_fig}
\let\spacing\relax
\usepackage{setspace}
\usepackage{lineno}
\usepackage{caption}
\usepackage{amsmath}
\usepackage[utf8]{inputenc}
\usepackage[T1]{fontenc}
\usepackage{sansmath,amssymb, amsmath, graphicx, grffile}
\usepackage{xcolor}
\usepackage{booktabs}
\DeclareCaptionLabelFormat{tablelabel}{Table~#2}

\usepackage{nameref} % To refer to Suppl. Methods by name

\usepackage[margin=1in]{geometry}
\usepackage{xspace}
\usepackage[colorlinks=true,citecolor=blue,linkcolor=blue,urlcolor=blue,breaklinks,hypertexnames=false]{hyperref} % ,draft

\newcommand{\apj}{The Astrophysical Journal}
\newcommand{\apjl}{The Astrophysical Journal Letters}
\newcommand{\aapr}{Astronomy \& Astrophysics Reviews}
\newcommand{\apjs}{The Astrophysical Journal Supplement Series}

\newcommand{\nat}{Nature} 
\newcommand{\aap}{Astronomy and Astrophysics}
\newcommand{\araa}{Annual Review of Astron and Astrophys}

\newcommand{\mnras}{Mon. Not. R. Astron. Soc.}   % Monthly Notices of the RAS
\newcommand{\prd}{Phys. Rev. D}   % Physical Review D
\newcommand{\pasp}{Publ. Astron. Soc. Pac.}   % Publications of the Astron. Soc. of the Pacific
\newcommand{\figmval}{$f_{\rm IGM}=0.76_{-0.11}^{+0.10}$ }
\newcommand{\fXval}{$f_{X}=0.15_{-0.10}^{+0.11}$}
\newcommand{\dmhost}{$130_{-23}^{+25}$}
\newcommand{\muhost}{$\mu_{host}=4.90_{-0.20}^{+0.18}$}
\newcommand{\sigmahost}{$\mathrm{\sigma_{host}} = 0.53_{-0.14}^{+0.16}$}
\newcommand{\fXDSA}{$f_{X} = 0.20_{-0.13}^{+0.13}$}
\newcommand{\figmDSA}{$f_{\mathrm{IGM}} = 0.70_{-0.13}^{+0.13}$}
\newcommand{\muhostDSA}{$\mu_{host} = 4.97_{-0.24}^{+0.22}$}
\newcommand{\sigmahostDSA}{$\sigma_{host} =  0.47_{-0.19}^{0.21}$}
\newcommand{\fXlowz}{$f_{X} = 0.08_{-0.06}^{+0.10}$}
\newcommand{\figmlowz}{$f_{\mathrm{IGM}} = 0.75_{-0.08}^{+0.07}$}
\newcommand{\muhostlowz}{$\mu_{host} = 4.76_{-0.36}^{+0.27}$}
\newcommand{\sigmahostlowz}{$\sigma_{host} = 0.65_{-0.19}^{+0.23}$}
\newcommand{\fd}{$f_d = 0.94_{-0.05}^{+0.05}$}
\newcommand{\fddsaonly}{$f_d = 0.89_{-0.06}^{+0.06}$}
\newcommand{\omegabh}{$\Omega_b\,h_{70} = 0.051_{-0.006}^{+0.006}$}
\newcommand{\fgas}{$f_{gas}=0.35_{-0.25}^{+0.30}\frac{\Omega_b}{\Omega_M}$}

\newcommand{\frbkoyaanisqatsi}{20221027A\xspace}
\newcommand{\frbnihari}{20221219A\xspace}

\newcommand{\frbmifanshan}{20221029A\xspace}
\newcommand{\frbbubble}{20240215A\xspace}
\newcommand{\frbannie}{20240213A\xspace}
\newcommand{\frbada}{20220831A\xspace}
\newcommand{\frbgemechu}{20231220A\xspace}
\newcommand{\frbcasey}{20240229A\xspace}
\newcommand{\frbnikhil}{20240119A\xspace}

\newcommand{\frbjohndoe}{20230814A\xspace}
\newcommand{\frbpushkin}{20240123A\xspace}
\newcommand{\frbbruce}{20230521B\xspace}

\newcommand{\sw}[1]{\texttt{#1}}

\protect
\begin{document}

\renewcommand\baselinestretch{1.2}

\begin{center}
    {\Large \bf A gas-rich cosmic web revealed by the partitioning of the missing baryons}
\end{center}
\vspace{0.25cm}
\begin{center}
    {\large Authors: 
    Liam Connor$^{1,2*}$, 
    Vikram Ravi$^{2,3}$, 
    Kritti Sharma$^{2}$, 
    Stella Koch Ocker$^{2,4}$, 
    Jakob Faber$^{2,}$,
    Gregg Hallinan$^{2,3}$, 
    Charlie Harnach$^{2,3}$, 
    Greg Hellbourg$^{2,3}$, 
    Rick Hobbs$^{2,3}$,
    David Hodge$^{2,3}$, 
    Mark Hodges$^{2,3}$, 
    Nikita Kosogorov$^{2,3}$,
    James Lamb$^{2,3}$, 
    Casey Law$^{2,3}$, 
    Paul Rasmussen$^{2,3}$, 
    Myles Sherman$^{2}$, 
    Jean Somalwar$^{2}$, 
    Sander Weinreb$^{2}$, 
    David Woody$^{2,3}$,
    Ralf M. Konietzka$^{1}$
}
\end{center}

\vspace{0.3cm}

\noindent $^{1}$ Center for Astrophysics | Harvard $\&$ Smithsonian, Cambridge, MA 02138-1516, USA\\
$^{2}$ Cahill Center for Astronomy and Astrophysics, MC\,249-17, California Institute of Technology, Pasadena CA 91125, USA. \\
$^{3}$ Owens Valley Radio Observatory, California Institute of Technology, Big Pine CA 93513, USA. \\
$^{4}$ The Observatories of the Carnegie Institution for Science, Pasadena, CA 91101, USA.\\
$^{*}$ E-mail: liam.connor@cfa.harvard.edu \\

\newpage

\textbf{Approximately half of the Universe's dark matter resides in collapsed halos; significantly less than half of the baryonic matter (protons and neutrons) remains confined to halos. A small fraction
of baryons are in stars and the interstellar medium within galaxies. The lion's share are diffuse ($<$\,10$^{-3}$ cm$^{-3}$) and ionized (neutral fraction $<$\,10$^{-4}$), located in the intergalactic medium (IGM) and in 
the halos of galaxy clusters, groups, and galaxies. 
This diffuse ionized gas is notoriously difficult to measure, but has wide implications for galaxy formation, astrophysical feedback, 
and precision cosmology. 
Recently, the dispersion of extragalactic 
Fast Radio Bursts (FRBs) has been used to measure the total content of cosmic baryons. Here, we present a large cosmological sample of FRB sources localized to their host galaxies. 
We have robustly partitioned the missing baryons into the IGM, galaxy clusters, and galaxies, providing a late-Universe measurement of the total baryon density of \omegabh. Our results indicate efficient feedback processes that can deplete galaxy halos and enrich the IGM (\figmval), agreeing with the baryon-rich cosmic web scenario seen in cosmological simulations. 
Our results may reduce the ``$S_8$ tension'' in cosmology, as strong feedback leads to suppression of the matter power spectrum.
}

\newpage

The DSA-110 is 
an interferometer operating between 1.28--1.53\,GHz at Caltech's 
Owens Valley Radio Observatory (OVRO)\cite{ravi2022dsa-110}. It is the first 
radio telescope built with the express purpose of detecting 
and localizing FRBs to their host galaxies, which is critical both for using FRBs as a cosmological tool\cite{lorimer2007, macquart20} and unveiling their physical origin\cite{cordesreview,petroffreview}. Between January 2022 and March 2024, that array was continuously observing in a commissioning period with 48 core antennas and 
15 outrigger antennas for offline arcsecond localization. During this time 60 new FRBs were discovered of which 39 now have a spectroscopic host galaxy redshift. The host galaxy properties of a uniformly selected subset of these sources are presented in a companion work\cite{SharmaNature2024}, where the Optical/IR follow-up procedure is described in detail. We present 9 sources that are not in that sample\cite{SharmaNature2024}.
We present three new FRBs near or beyond redshift 1, which constrain the IGM column by virtue of their great distance.

\begin{figure}
\centering
\includegraphics[angle=0,width=1\textwidth]{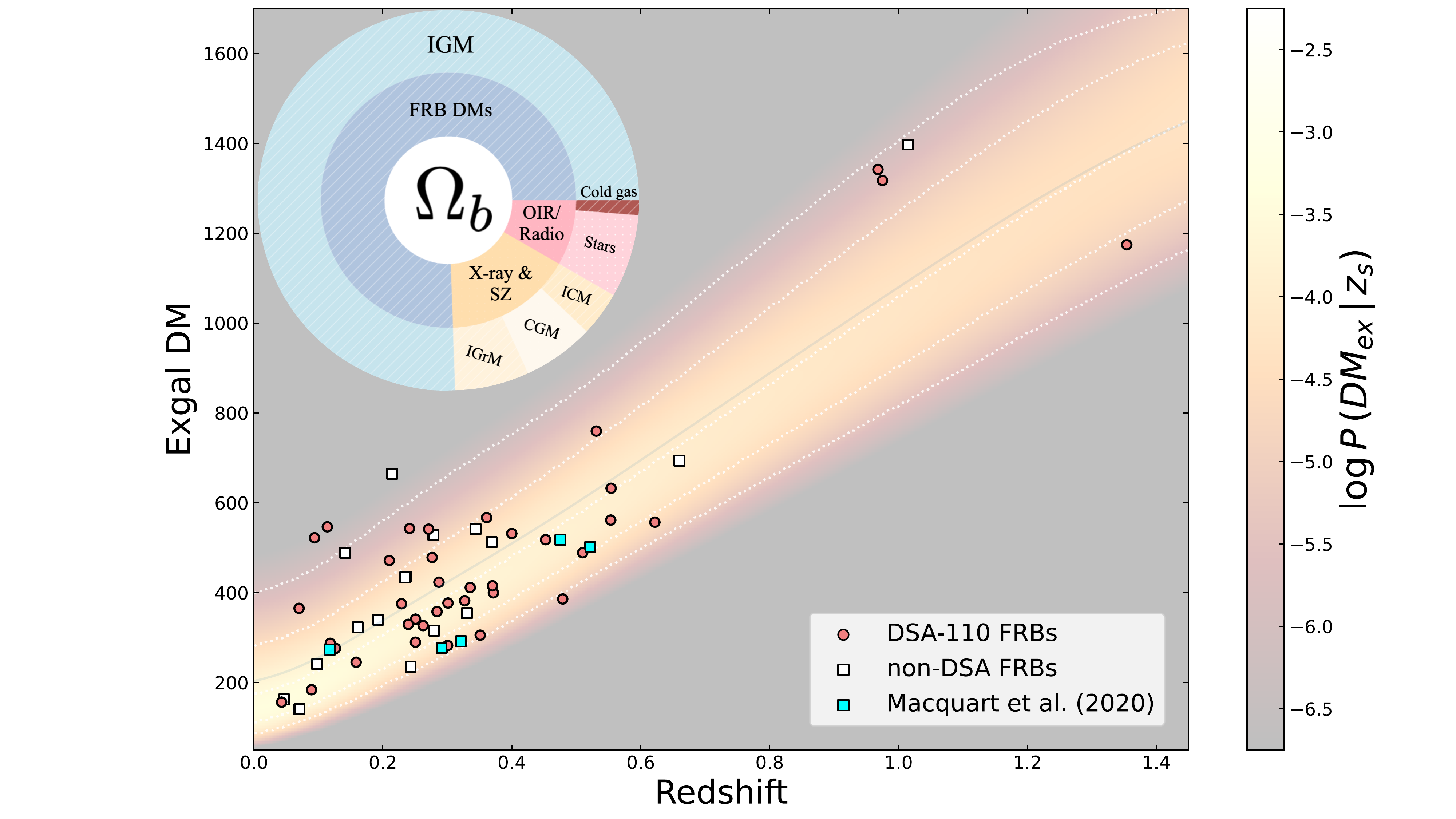}
\caption{{\bf A full account and partition of the missing baryons.} We show the distribution of extragalactic dispersion measure and redshift of 
localized FRBs in our sample, over half of which are sources recently discovered by the DSA-110. The heatmap corresponds to a likelihood function $P(\mathrm{DM}_{ex} | z_s)$ with $f_{\mathrm{IGM}}=0.80$ and $f_X=0.10$ and a 
log-normal host DM distribution. The five cyan squares are the sample that was used to derive the original 
``Macquart Relation''\cite{macquart20}. The solid dark curve is the mean DM at that $z_s$ and white curves show the median, 1-$\sigma$ and 2-$\sigma$ contours. 
The radial treemap 
inset figure shows our comprehensive partition 
of cosmic baryons. We have created this chart by synthesizing  
our FRB results with other precision probes of the Universe's normal matter, including the baryon budget of groups and clusters, cold gas, and stars as determined by various methods.
\label{fig:mainfig}}
\end{figure}

We append our sample to 30 previously localized FRB sources. 
The distribution of extragalactic DM and redshift for our 
full sample is plotted in Figure~\ref{fig:mainfig}.
Their positions, DM, redshifts, and detection instrument are displayed in Table~1. The 9 FRBs unique to this work are in Table~2 and their host galaxy image mosaic is shown in Extended Data Figure 1. The extragalactic DM of localized FRBs encodes information about the quantity and distribution of diffuse baryons in the Universe. The total observed DM of an FRB can be written as the 
sum of several components along the line of sight,

\begin{equation}
   \mathrm{DM_{obs} = DM_{MW} + DM_{IGM}}(z_s) + \sum_{i}^{N_X}\,\frac{\mathrm{DM_{X}(M_i}, b_\perp)}{1+z_i} + \frac{\mathrm{DM_{host}}}{1+z_s}.
\end{equation}

\noindent In $\rm DM_{MW}$ we include both the Milky Way's interstellar medium (ISM) and halo. $\rm DM_{IGM}$ corresponds to the ionized gas outside of virialized halos and in the intergalactic medium\cite{mcquinnIGM15}, and $\mathrm{DM_{X}(M_i}, b_\perp)$ is the rest-frame contribution from the $i^{th}$ intersected halo. The latter value will depend on the halo mass $\mathrm{M}_i$, redshift $z_i$, and the physical impact parameter $b_\perp$ (higher masses and smaller offsets lead to more DM). $\rm DM_{host}$ is from gas in the FRB host galaxy and may come from the halo, its ISM, and/or circumsource plasma. $\rm DM_{IGM}$ is expected to dominate for sources beyond $z\approx0.2$, unless the sightline intersects a galaxy cluster\cite{connor23} or the FRB is embedded in an unusually dense local environment\cite{niu2022}. If we define a ``cosmological DM'' as $\rm DM_{cos} \equiv DM_{IGM} + DM_X$, then the average sightline's DM from the IGM and intervening halos is,

\begin{equation}
    \left < \mathrm{DM_{cos}} \right > = \frac{3c\,\Omega_b\,H_0}{8\pi\,G\,m_p} \int_0^{z_s} \frac{(1+z)\,f_d(z)\,f_e(z)}{\sqrt{\Omega_{\Lambda} + \Omega_m (1+z)^3}}\,\mathrm{d}z.
\label{eq:dmcos}
\end{equation}

\noindent where $\Omega_b$ is the cosmic baryon abundance, 
$\Omega_m$ is the matter density parameter, $\Omega_\Lambda$ is the dark energy parameter, $H_0$ is the Hubble constant, $m_p$ is the proton mass, $f_e$ is the number of free electrons per baryon, and $f_d$ is the baryon 
fraction in the diffuse ionized state (i.e. not in stars or cold neutral gas). Taking $f_e=0.875$ and $f_d(z)$ to be constant and then using $h_{70}\equiv H_0 / (70\,\rm km\,s^{-1}\,Mpc^{-1})$, one finds  $\left < \mathrm{DM_{cos}} \right > \approx 1085\,z\, f_d\,\left (\frac{\Omega_b\,h_{70}}{0.04703} \right)\,\,\,\mathrm{pc\,cm^{-3}}$ for $z_s \lesssim1$. 

Our FRB analysis is centered on the extragalactic DM distribution as a function of redshift, $z_s$, where $\rm DM_{ex} = DM_{obs} - DM_{MW}$.
For each source in our sample, we compute a 1D likelihood function 
$P(\mathrm{DM_{ex}} | z_s, \vec{\theta})$ where the model parameters are 
$\vec{\theta}=\left \{ f_{\mathrm{IGM}}, f_X, \mu_{host}, \sigma_{host} \right \}$. 
Our new method explicitly parameterises
the fraction of baryons in the IGM, $f_{\mathrm{IGM}}\equiv\frac{\Omega_{\mathrm{IGM}}}{\Omega_{b}}$, and in intersected halos, $f_{X}\equiv\frac{\Omega_{halos}}{\Omega_{b}}$ referenced to redshift 0.1 (see Methods). The three components of gas, $\rm DM_{\rm IGM}$, $\mathrm{DM}_X$, and $\mathrm{DM}_{host}$, are in principle separable for a sufficiently large sample
because each has a different redshift dependence and  
$P(\mathrm{DM} | z_s)$ distribution.
Our effective definition of the IGM is gas outside of 
virialized dark matter halos. 
From the per-source likelihoods, we compute a 
posterior over all FRBs as $\prod\limits_{i}^{N_{FRB}} P(\mathrm{DM_{t, i}} | z_{s,i})P(\vec{\theta})$ which we estimate using Markov chain Monte Carlo (MCMC). 
We assume a 
log-normal distribution for the host contribution to DM with 
parameters $\mu_{host}$ and $\sigma_{host}$\cite{macquart20}, which are the log-normal mean and standard deviation, respectively. We take a wide, flat prior on the log-normal mean, $p(\mu_{host}) \sim \text{Uniform}(0, 7)$, allowing the median host DM to span 0 to 1,000\,pc\,cm$^{-3}$. We assume the same flat prior on both $f_{\mathrm{IGM}}$ and $f_{X}$ of $\text{Uniform}(0, 1)$ with the added constraint that $f_{\mathrm{IGM}} + f_{X} \leq 1$. We take a simulation-based 
inference approach as our primary method of fitting cosmic gas parameters, using 
a mock FRB survey in IllustrisTNG\cite{walker2023} as a baseline. 
Large hydrodynamical simulations are valuable for this task 
because of the complex relationship between the dark matter distribution, galaxy formation, and baryons, which cannot be described analytically.

\begin{figure*}
\centering
\includegraphics[width=0.99\linewidth]{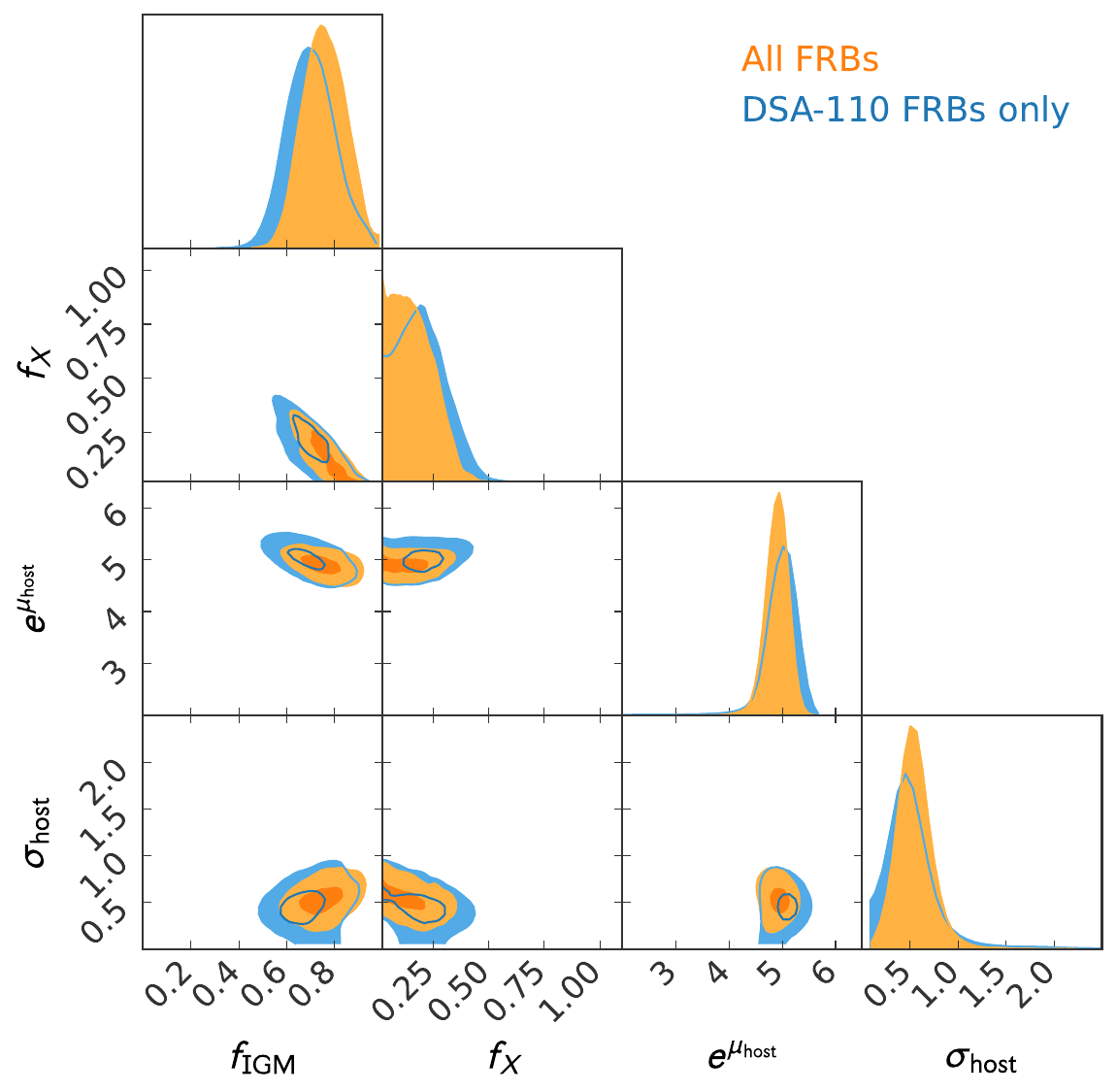}   
\caption{{\bf An MCMC fit of extragalactic gas parameters to 
a sample of localized FRBs.} We show the joint posteriors 
of the fraction of baryons in the IGM, $f_{\rm IGM}$, 
in halos, $f_X$, as well as those of a log-normal 
distribution taken to describe the host galaxy DM, $\mu_{host}$
and $\sigma_{host}$. The orange regions correspond to the 
results of fitting the full FRB sample; blue is a sub-sample 
of only DSA-110 discovered sources, which produces 
consistent values with slightly larger uncertainty. The 
shaded regions are 1-$\sigma$ and 2-$\sigma$ contours. 
Previously, $f_{\rm IGM}$ was highly uncertain because intergalactic gas is hot, diffuse, and difficult to detect directly. Our results demonstrate that four fifths of all
baryons occupy the cosmic web, outside of dark matter halos.
\label{fig:corner}}
\end{figure*}

\begin{figure*}
\includegraphics[width=0.98\textwidth]{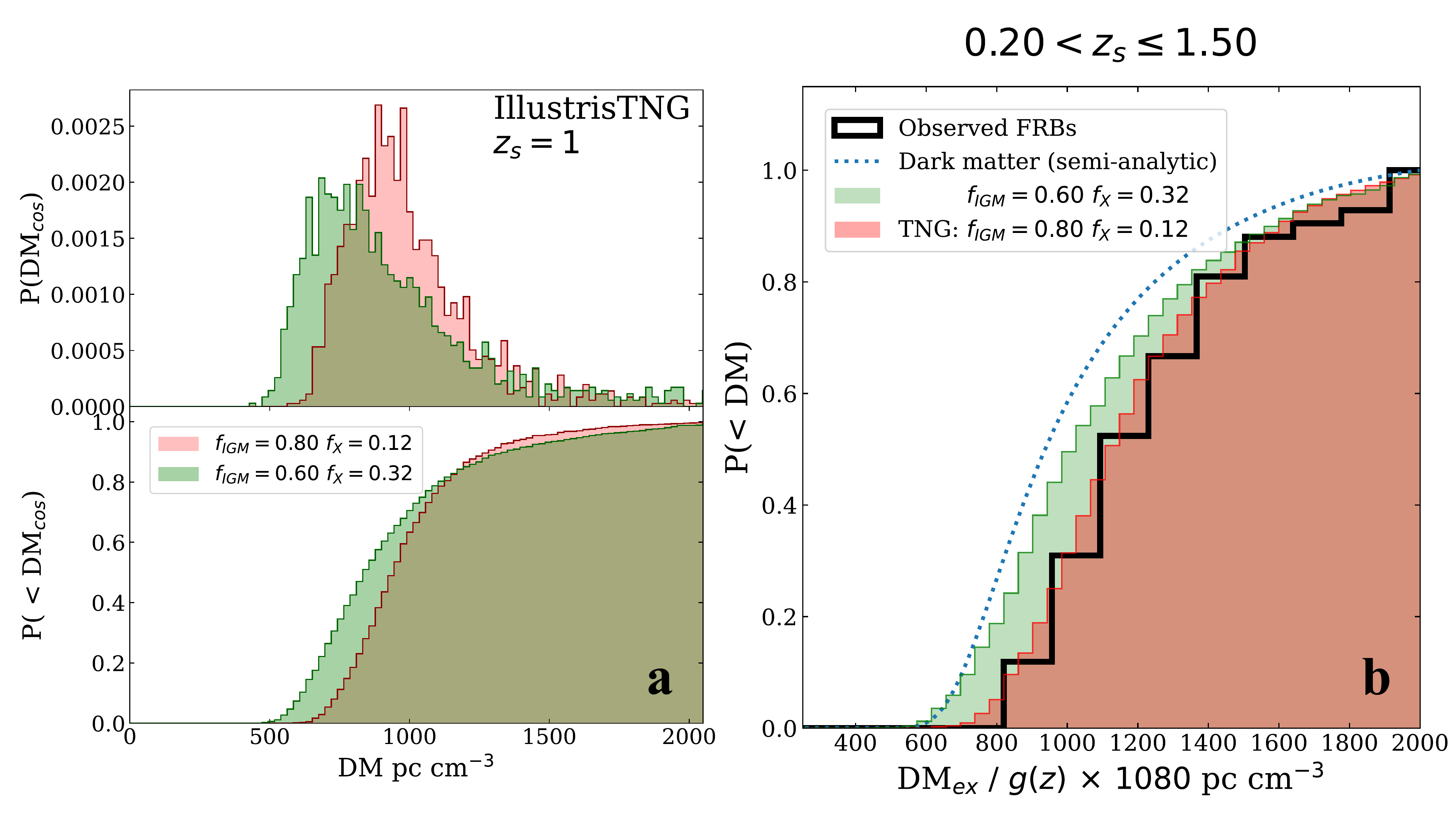}
\caption{{\bf The extragalactic DM distribution of FRBs prefers a gas-rich intergalactic medium}. In panel $\mathbf{a}$
we show the simulated cosmic DM distributions (i.e. beyond the Milky Way but excluding the host DM) for two scenarios. Red histograms represent a gas-rich 
IGM with $f_{\rm IGM}=0.80$ 
and baryon-deficient halos where 
$f_X=0.12$; 
green histograms 
correspond to $f_{\rm IGM}=0.60$ and 
$f_{X}=0.32$. 
Panel $\mathbf{b}$ shows the respective cumulative DM distributions. The green and red distribution have the same mean cosmic DM, despite 
very different shapes at below-median DM values.
The blue dotted curve is a semi-analytic DM distribution assuming the baryons perfectly trace dark matter. As expected from the MCMC fits, the scenario with 
low $f_{\rm IGM}$ 
and baryon-rich halos is disfavoured by our 
data, as is the scenario where baryons trace dark matter. 
\label{fig:pdmsims}}
\end{figure*}

The fit to our primary dataset of all eligible FRBs produces \figmval and \fXval, 
as shown in the corner plot in Figure~\ref{fig:corner}. The 
large value of $f_{\rm IGM}$ emerges from a strong feature 
in our data: The lack of FRBs with low
values of extragalactic DM per redshift, sometimes
referred to as the ``DM Cliff''\cite{jamesH02022} (more detail is provided in Methods). This implies a smooth Universe and a significant
minimum DM value from the IGM. For example, none of our 
sources beyond redshift 0.1 has $\frac{\mathrm{DM_{ex}}}{z_s} < 800\,\, $\,pc\,cm$^{-3}$. 
If extragalactic DM were dominated by intervening halos or the host galaxies, 
we would expect a less pronounced rise in $P(\mathrm{DM_{ex}} | z_s)$ 
because most sightlines in our sample do not intersect a halo. The same is true for a Universe in which baryons trace perfectly the dark matter, as can be seen in
the blue dotted curve of Figure~\ref{fig:pdmsims}. Instead, 
the IGM provides a significant statistical floor in DM per unit 
distance because most sightlines 
intersect many dozens of filaments\cite{walker2023} and even cosmic voids contribute a considerable electron column. Our FRB sample rules out scenarios where baryons trace dark matter, in which $f_{\rm IGM}$ is low and a large 
portion of the missing baryons are confined to galaxy halos.
We infer \muhost and \sigmahost. 
This corresponds to a modest median rest-frame host DM contribution of \dmhost\,pc\,cm$^{-3}$ for this sample.
We fit sub-samples of the data, for example DSA-110-detected sources only. We have also performed jackknife resampling excluding/including individual sources (high-$z_s$, re-introducing sources with large excess DM, etc.). While the sources at $z_s\gtrsim0.5$ have significant constraining power, in all cases our data prefer a large fraction of baryonic material in the IGM and a large total diffuse fraction, $f_d$.

Although FRB DMs are impacted by the ionized gas in galaxy groups and clusters\cite{Prochaska19, connorravi22, connor23}, the most precise constraints on the baryon budget in massive halos
come from X-ray\cite{Sun2009, eRASS1-clusters2024} and 
Sunyaev–Zeldovich (SZ) measurements\cite{psz2,act_clusters}. Thermal X-ray emission is $\propto \!\int n_e^2\,dl$ and 
SZ is $\propto\!\int n_e\,T_e\,dl$, where $n_e$ and $T_e$ are the free electron density and temperature respectively, so both are sensitive to 
large, dense regions of hot gas. In contrast, FRBs pick up DM from all ionized plasma along the line of sight. The hot baryon fraction in halos, $f_{hot}$, is a function of halo mass, approaching the cosmological ratio $\approx\frac{\Omega_b}{\Omega_M}$ for the most massive galaxy clusters\cite{Gonzalez_2013}.
This quantity is less certain 
for halos below $10^{14}\,h^{-1}_{70}\,M_\odot$\cite{DaiBregman2009}. However, recent advances 
in sample sizes and measurement
precision\cite{vikram2017, eRASS1-clusters2024} 
have significantly improved our knowledge of the cluster mass 
function and $f_{hot}$. We have synthesized these multiwavelength 
observations to estimate 
the fraction of the Universe's baryons in 
the hot gas of galaxy groups and clusters.
We find $f_{\rm ICM} = 3.75\,\pm 0.5\,\%$ of all baryons are in the intracluster medium (ICM). For galaxy groups with 
$10^{12.7}\,M_\odot \leq M_h \leq 10^{14}\,M_\odot$ this number is $5.4\,\pm 1.0\,\%$. Together, we conclude that 
roughly $9\%$ of baryons are in a diffuse ionized state in massive halos.

Next, we consolidate estimates of the baryon fraction in galaxies, including stellar mass and cold gas. These are the last major components of the baryon budget. 
The majority of cold gas in the Universe is 
neutral atomic hydrogen, with traces of molecular hydrogen and helium.
At low redshifts, 21\,cm galaxy surveys measure the HI mass function which can then be integrated to estimate the neutral hydrogen density\cite{Zwaan2005}. 
We take the values for $\rm \Omega_{HI}$, $\rm \Omega_{H_2}$, and their associated uncertainty from a recently-assembled suite of volumetric surveys\cite{walter20}, finding 
$f_{\rm HI}=9.6_{-2.3}^{+3.8}\times10^{-3}$, $f_{\rm H_2}=1.6_{-0.4}^{+0.8}\times10^{-3}$ and $f_{cold}=1.1_{-0.2}^{+0.3}\times10^{-2}$. Thus, just over one percent of the Universe's baryons are in cold neutral gas within galaxies. The total baryon content of stars and stellar remnants is larger, but more difficult to model\cite{fukugita1998, madaudickinson2014}. Most stellar mass is in low-mass stars, so the baryon fraction in stars and stellar remnants, $f_*$, is sensitive to the chosen initial mass function (IMF) which dictates the number 
of low-mass stars. Using the bottom heavy Salpeter IMF\cite{Salpeter1955} (i.e. many low-mass stars), $f_*$ can be as high as 14$\%$ at low redshifts\cite{Gallazzi2008,madaudickinson2014}. If instead we opt for a Chabrier IMF (fewer low-mass stars) and use a smooth fit to multiple measurements of $\rho_*(z)$, we find $f_*\approx4-7\%$\cite{walter20,Leja2020} (see Methods for a detailed discussion).

We fully account for the so-called missing baryons.
More importantly, we are able to partition them into the IGM, galaxy groups, galaxy clusters, and galaxies after synthesizing our FRB results with other observations. A significant majority of baryonic matter resides in the IGM, outside of virialized halos. From our FRB-independent analysis of X-ray groups and clusters, we find that $9.2_{-1.6}^{+1.6}\%$ are in an ionized phase occupying massive halos. Roughly one percent are in cold neutral gas in galaxies. This leads 
us to conclude that the CGM of individual galaxies cannot harbor a substantial fraction of the baryons in the Universe. The global analysis is 
in agreement with detailed studies of individual FRB source sightlines. Our FRBs 
that intersect one or more foreground galaxy CGM at low impact parameter do not have significant excess dispersion (see Methods). 
We find that \fgas  for $10^9\,M_\odot < M < 5\times10^{12}\,M_\odot$, 
below the cosmic average. These results require feedback processes\cite{TumlinsonCGM} 
to expel and/or prevent gas from falling into their potential wells. We cannot discriminate between specific models, but our picture of a rich IGM and baryon-deficient CGM 
is consistent with simulations where feedback suppresses lower-mass baryon halos \cite{Sorini2022,TNG50_CGM2022}. 
For example, with all feedback turned off 
in the SIMBA simulation\cite{Dave2019} it was found that 
$f_{\rm IGM}\approx0.6$ by $z<1$; with AGN 
feedback turned on, $f_{\rm IGM}$ was over eighty five percent\cite{Sorini2022}. Similarly, in 
IllustrisTNG $f_{\rm IGM}\approx80\%$ at 
low redshifts\cite{walker2023, artale2022} and baryons 
are missing from the CGM of Milky Way-like galaxies\cite{TNG50_CGM2022} (see Extended Data Figure~2). Our findings also agree with recent statistical cross-correlations of galaxy surveys with X-ray\cite{hotCGMerass2024} and kinematic Sunyaev-Zel’dovich (kSZ)\cite{ksz2024}, indicating a dearth of baryons confined to galaxy halos. Our findings may alleviate some of the ``$S_8$ tension'' in cosmology\cite{sigma8}, 
where weak lensing surveys have reported lower fluctuation amplitudes 
in the large-scale structure than \textit{Planck's} best fit $\Lambda$CDM parameters. If gas is evacuated from halos into the IGM by strong feedback, 
observed weak lensing signals will be smaller than expected\cite{amonsigma8}.

\begin{figure*}
\includegraphics[width=\textwidth]{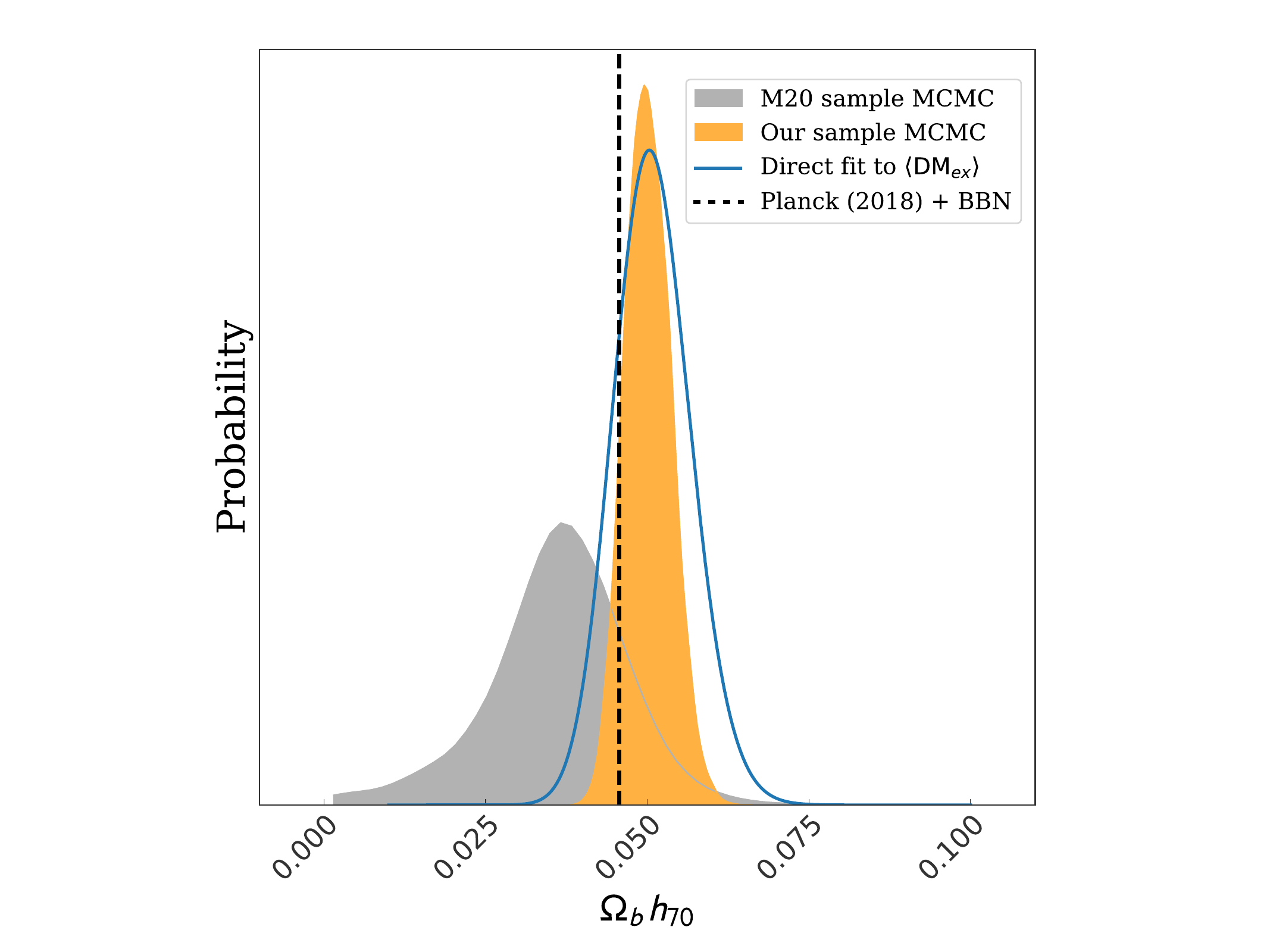}
\caption{{\bf A ten percent measurement of the present day baryon parameter}: We show constraints on $\Omega_b\,h_{70}$ for our sample of FRBs. 
These fits are independent of the partition of hot gas into the IGM and halos because cosmic DM is proportional to 
$\Omega_b\,h_{70}$, assuming a prior on $f_d$.
Our results agree with cosmic microwave background (CMB) and Big Bang Nucleosynthesis (BBN) measurements with unprecedented precision, bridging the gap between the early and late Universe.
We fit our data with two methods. The first is an MCMC over all individual FRBs which fits for the host 
DM distribution as well as $\Omega_b\,h_{70}$ (orange). The second is a maximum-likelihood estimate on the mean cosmic DM (blue). 
We also fit our model to a previous sample (M20, Macquart et al. (2020) in grey), which was the first to use FRBs to constrain the baryon density parameter.}
\label{fig:omega_b_fig}
\end{figure*}

We have attempted to construct a model with enough flexibility 
that any physically realistic partition of cosmic gas could be fit 
(see Extended Data Figure~3). 
However, 
our primary simulation-based inference approach uses a single 
cosmological simulation, TNG300, adding uncertainty to our fit.  
We account for this with a $30\%$ increase of our MCMC errors based 
on model mismatch resampling experiments.
To verify the large derived value of $f_{\rm IGM}$, we apply an independent technique that directly quantifies the total diffuse gas fraction, $f_d$, using the
average cosmic dispersion of our sample, $\left < \rm DM_{cos}(z_s) \right > $ (see Equation~\ref{eq:dmcos}). This method does not rely on a partition of the cosmic baryons, as the quantity is sensitive to both intergalactic and intervening halo gas. 
We find \fd, independent of any assumptions about $P_{cos}(\mathrm{DM_{IGM}}, \mathrm{DM}_X)$ and its redshift evolution (see Extended Data Figure~4). We have also tested 
our large $f_{\rm IGM}$ result with a semi-analytic DM distribution in which 
baryons trace dark matter, and 
by a post-hoc alteration to the IllustrisTNG mock FRB survey (see Methods).
If we instead make the Universe's total cosmic baryon content a free parameter, we find \omegabh \,(see Figure~\ref{fig:omega_b_fig}). This late-time measurement is consistent at the 10$\%$ level with early Universe constraints of the physical baryon density from Big Bang Nucleosynthesis\cite{Cooke_2018} and the Cosmic Microwave Background\cite{Planck2018}. An equally precise constraint can be obtained for the Hubble constant, resulting in $H_0=71^{+7}_{-7}$\,\,km\,s$^{-1}$\,Mpc$^{-1}$. This comes with an important caveat that disentangling $H_0$ from $\Omega_b$ requires fixing the baryon density parameter at the early Universe value, rendering a direct and independent measurement of $H_0$ with FRBs difficult \cite{Wu_2022, Hagstotz_2022}. 

Notably, the mean cosmological DM of
FRBs places a ceiling on the total stellar mass in the Universe because $f_* < 1 - f_d - f_{cold}$, for a given $\Omega_b\,h_{70}$. 
Our results suggest that over ninety percent of baryons are in the diffuse ionized state or in cold gas (i.e. not in stars). This constraint is independent of galaxy spectral energy distribution modeling, choice of the IMF, and the low mass cut-off that affects typical methods\cite{madaudickinson2014}. Since 
most stellar mass is bound in the abundant low-mass stars, 
our $f_*$ upper-limit constrains the mean stellar IMF. We place a $90\%$ upper 
limit on the stellar baryon fraction at low redshifts of $f_*\leq9\%$ and therefore $\rho_* \leq 5.6\times10^{8}\,M_\odot\,\mathrm{Mpc}^{-3}$.
We use this to rule out Salpeter IMFs for a low-mass cut-off below $0.10\,M_\odot$.

\newpage

\newpage

\clearpage

\newpage

\clearpage

\begin{center}
    {\Large \bf Methods}
\end{center}
\vspace{0.5cm}

\noindent {\bf New DSA-110 fast radio bursts}
In addition to the sample described in our companion paper~\cite{SharmaNature2024}, we present nine new FRBs discovered with the DSA-110 between February 2022 and February 2024. All FRBs were discovered and localized using techniques summarized in Sharma et al. (2024)~\cite{SharmaNature2024}. The nine additional FRBs were not included in the companion paper because they did not meet either the host-galaxy magnitude selection criterion, or the date cutoff. 

% \textbf{very briefly describe optical follow-up, referring to new figures and tables}

Out of these nine FRBs, four have a candidate host galaxy detected in archival $r$-band from the Beijing-Arizona Sky Survey (BASS) from the Dark Energy Survey~\cite{2017PASP..129f4101Z} and two have a candidate host galaxy detected in archival $r$-band data from PanSTARRS1 (PS1)~\cite{2016arXiv161205560C}. For the other four candidates, we obtain deep optical/IR imaging with the Low Resolution Imaging Spectrometer (LRIS)~\cite{1995PASP..107..375O} on Keck-I, DEep Imaging Multi-Object Spectrograph (DEIMOS)~\cite{2003SPIE.4841.1657F} on Keck-II at W. M. Keck Observatory and the Wide Field Infrared Camera (WIRC)~\cite{2003SPIE.4841..451W} instrument, mounted on the 200-inch Hale Telescope at the Palomar Observatory. These data were reduced either using standard procedures (such as \sw{LPipe}~\cite{2019PASP..131h4503P} for LRIS) or a custom pipeline, as described in our companion paper. 

Having obtained plausible host candidates for all host galaxies, we use the Bayesian Probabilistic Association of Transients to their Hosts (PATH) formalism~\cite{2021ApJ...911...95A} to estimate the host association probability (P$_{\mathrm{host}}$). We find that all of these FRBs have secure host galaxy associations with P$_{\mathrm{host}} > 95$\%. The imaging mosaic of the host galaxies of these FRBs is shown in Extended Data Figure~1 and the basic FRB properties are listed in Table~2. Next, we obtain their optical/IR spectroscopy with Keck:I/LRIS, Multi-Object Spectrometer For Infra-Red Exploration (MOSFIRE)~\cite{2012SPIE.8446E..0JM}
at the Keck Observatory and Double Spectrograph (DBSP)~\cite{1982PASP...94..586O} on the 200-inch Telescope at Palomar Observatory. The spectroscopy setups used are summarized in Supplementary Table~1. These data were reduced using the \sw{LPipe}~\cite{2019PASP..131h4503P}, the Python Spectroscopic Data Reduction Pipeline (\sw{PypeIt})~\cite{pypeit:zenodov_v1_6, pypeit:joss_pub}, and the \sw{DBSP\_DRP}~\cite{dbsp_drp:joss} software. The 1D spectra for bright hosts and 2D spectra for faint hosts are shown in Supplementary Figure~1 and 2 respectively.

The redshift we adopt for \frbbruce is based on the detection of a single line in its MOSFIRE spectrum. The line does not overlap any night-sky emission lines, was not intersected by any cosmic ray hits, and is evident in jack-knife subsets of exposures. We assume that this line corresponds to H${\rm \alpha}$. If this line corresponds to any other bright line in a typical galaxy spectrum\cite{Kewley2019}, such as [OIII]$\lambda$5007 and [OII]$\lambda$3727, \frbbruce would be at an unrealistically high redshift. We also note that a deep optical spectrum of \frbbruce obtained with LRIS, which attained approximately $3\sigma$ sensitivity per resolution element of $7\times10^{-19} \rm erg s^{-1} cm^{-2} A^{-1}$, 
did not reveal any emission lines. 

Since our goal 
is to measure the cosmological distribution of ionized gas, we treat the $\rm DM_{host}$ and its variance as nuisance parameters to be marginalized over. We are free to exclude FRBs for which there is evidence of strong local dispersion. If such sources are correctly identified and are not correlated with the large-scale structure, their removal should not bias our inference of the IGM and halo gas parameters. We exclude 
two sources for this reason, FRB\,20190520B\cite{niu2022} and FRB\,\frbada\cite{caseyAda2024}, which both have strong evidence for excess local dispersion (large Faraday rotation measure, scattering, which are not caused by the IGM). 

\noindent {\bf The full localized FRB sample} 
Over half of the FRBs in our sample were detected and localized by the DSA-110\cite{ravi2022dsa-110}. 
The majority of non-DSA FRB localizations have come from 
the ASKAP CRAFT survey\cite{CRAFT2017,prochaska2019}, with a handful of others from CHIME/FRB\cite{M81mohit, M81Kirst}, 
MeerKAT\cite{caleb2023}, and realfast on the VLA\cite{clawrealfast}. 

We estimate $\rm DM_{ex}$ for each FRB as DM$_{obs}$ minus the Milky Way DM, which 
we take to be the NE2001 value\cite{ne2001} in that direction plus the halo contribution. We exclude from our sample sources whose modeled Milky Way DM is greater than 40$\%$ of the total observed DM, since the uncertainty in Galactic ISM models is difficult to constrain. This
effectively discards sources that are both nearby and at low Galactic latitude.
Most of our sources lie at high Galactic latitudes, where the ISM contribution to $\rm DM_{MW}$ is dominated by the thick disk. The maximum DM contribution of the thick disk is well-constrained empirically by pulsars with measured distances (especially pulsars in distant globular clusters), the most recent constraint giving ${\rm DM_{max}} = (23.5 \pm 2.5\ {\rm pc\ cm^{-3}}) \times {\rm csc}|b|$ where $b$ is Galactic latitude\cite{ocker2020}. This maximum DM estimate is entirely consistent with NE2001 and has $\approx 10\%$ uncertainties for latitudes $\gtrsim 10^\circ$ (barring the presence of discrete structures along the LOS, which are rare at high latitudes and likely irrelevant for the FRBs we consider)\cite{ocker2020}. We adopt a constant value 
for the Milky Way halo DM of 30\,pc\,cm$^{-3}$. This is lower than previous estimates due to recent evidence from nearby extragalactic FRBs and globular cluster pulsars 
that the Milky Way CGM contributes a modest amount 
of DM\cite{ravi2022mark, cookfrb2023}. However, 
we have run our MCMC fits assuming the halo DM is 
50\,pc\,cm$^{-3}$ and 10\,pc\,cm$^{-3}$. In both cases 
the cosmic gas parameters are unaffected at the $\sim$\,percent level, but DM is shifted from/to the host galaxy DM parameters.

We include in our analysis only FRBs for which there is a 
robust host-galaxy association. Below we describe several notable sources, including those that we have chosen to exclude from our sample. 

\noindent\textit{FRB\,\frbbruce}: This is the highest published 
redshift of any FRB to date. 
We find no evidence for significant excess DM due to foreground structures. Instead, the 
extragalactic DM is slightly below median for this redshift, providing 
significant constraining power on $f_{\mathrm{IGM}}$ thanks to its high redshift
and proximity to the ``DM Cliff''. In Supplementary Figure~2 we show its host galaxy spectrum.

\noindent\textit{FRB\,\frbmifanshan}: This source is at redshift 0.975 with $\rm DM_{ex} \approx 1350$\,pc\,cm$^{-3}$. It passes by the outskirts of one massive galaxy cluster ($10^{14.8}\,M_\odot$ at $b_\perp\approx1.28$\,Mpc, cluster J092758+721851\cite{yang2021}) and passes through at least one massive galaxy group (J092809+722355). Still, it is unlikely that the ICM or IGrM dominates the total extragalactic DM given the high impact parameters. The correlation of filamentary structure 
with massive halos means the IGM along this line of site probably contributes 
excess DM compared with other sightlines for $z\approx1$.

\noindent\textit{FRB\,\frbpushkin}: The host galaxy is at redshift 0.968 with $\rm DM_{ex} \approx 1370$\,pc\,cm$^{-3}$. Notably FRB\,\frbpushkin passes within 3.9\,arcminutes of the cataloged position of the nearby ($\sim3$\,Mpc) dwarf spiral galaxy NC\,1560. Indeed, the burst is observed through the visible stellar disk of the galaxy. It is unlikely that the galaxy, which has a stellar mass of $\sim5\times10^{8}M_{\odot}$\cite{Gentile2010}, contributes significantly to the DM budget of FRB\,\frbpushkin.

\noindent\textit{FRB\,\frbnihari}: The sightline of FRB\,\frbnihari
is crowded, despite its limited
extragalactic dispersion\cite{faber2024} ($\rm DM_{ex}=662$\,pc\,cm$^{-3}$ at $z_s=0.554$).
The FRB passes through a $z\approx0.14$ galaxy cluster with mass $\sim$\,10$^{14.11}\,M_\odot$ (cluster J171039.6+713427) at its virial radius. 
The ICM appears not to contribute a large amount of DM, which could be due to the significant variation within and 
between clusters in free electron column at a fixed
$b_\perp$\cite{connor23}.
The pulse also traversed the CGM of two foreground galaxies behind the cluster but in front of the host
($b_\perp \approx 43.0_{-11.3}^{+11.3}$\,kpc for a $M_* = 10.60_{-0.02}^{+0.02}\,M_\odot$ galaxy and $b_\perp \approx 36.1_{-11.3}^{+11.3}$\,kpc for a $M_* = 10.01_{-0.02}^{+0.02}\,M_\odot$ galaxy). The
tight $\rm DM_{ex}$ budget constrains $\rm DM_{CGM}$ to be less than $\sim$\,40\,pc\,cm$^{-3}$ for both galaxies combined, bolstering our finding that a significant fraction of the Universe's baryons are in the IGM, not in galaxy halos.

\noindent\textit{FRB\,\frbgemechu}: Similar to FRB\,\frbnihari, this 
source passes through the CGM of two foreground galaxies. 
The sightline traverses the halo of NGC\,2523 at $b_\perp \approx 70$\,kpc
and comes within $\sim$\,30\,kpc of the barred spiral 
galaxy UGC\,4279. The burst's extragalactic DM was 
$\sim$\,410\,pc\,cm$^{-3}$ with a host galaxy redshift of 0.3355, indicating limited excess DM from the foreground halos.

\noindent\textit{FRB\,20190520B}: This repeating FRB source is highly scattered\cite{ocker2023a}, 
has a large and rapidly varying Faraday rotation measure (RM)\cite{reshamRM2023}, and originates
in a star-forming dwarf galaxy at $z=0.241$\cite{niu2022}. 
Its excess DM is $\mathcal{O}(10^3)$\,pc\,cm$^{-3}$. Thus, there is good evidence that FRB\,20190520B is dispersed in nearby plasma as well as the host galaxy's ISM. It also appears to be impacted by at least one foreground galaxy cluster\cite{kglee23}, but the fast dynamics demand an active local environment. We elect to exclude FRB\,20190520B from our primary MCMC 
fits because of the large local DM. This biases our estimates of nuissance parameters $\mu_{host}$ and $\sigma_{host}$ but only removes noise from our estimate of $f_{\rm IGM}$ 
and $f_X$. However, we have reintroduced this source to our fits during jackknife tests.

\noindent\textit{FRB\,\frbada}: With a low-mass host galaxy at $z_s=0.2620$,
no significant foreground structure, and roughly four times more extragalactic DM
than typical sightlines at the same redshift, we conclude that FRB \frbada has a large 
$\rm DM_{host}$. The sightline was cross-matched against several 
galaxy group and cluster catalogs\cite{caseyAda2024} with no large foreground halos within 5\,Mpc in transverse projection. We exclude it from our cosmic sample.

\noindent\textit{FRB\,20190611B}: We follow previous works\cite{macquart20} by excluding FRB\,20190611B from our analysis, 
as its host galaxy association is not secure: There are several galaxies near the source with ``chance-coincidence'' probabilities above 0.9\cite{PATHpaper}. Its PATH 
probability is also too low to be included in our sample.

\noindent\textit{FRB\,\frbkoyaanisqatsi}: We elect to 
exclude this FRB from our sample because of its ambiguous host galaxy. Two candidate galaxies at $z=0.5422$ and 
$z=0.2290$ could each plausibly be the host (in the absence of a strong prior on DM($z_s$)).

\noindent {\bf A direct estimate of $\mathbf{f_d}$} The average cosmic DM  
(i.e. $\rm DM_{IGM} + DM_X$) of a large sample of localized FRBs is a proxy 
for the total ionized baryon content of the Universe, $f_d\,\Omega_b$, 
independent of the partition between halos and the IGM.
Taking $f_d$ to be roughly constant for $z\lesssim1.5$,

\begin{equation}
    \left < \mathrm{DM_{cos}} \right > \approx  f_d\,K\,\Omega_b\,h_{70} \int_0^{z_s} \frac{(1+z)\,\mathrm{d}z}{\sqrt{\Omega_{\Lambda} + \Omega_m (1+z)^3}},
\label{dmcos}
\end{equation}

\noindent where $K=\frac{3c}{8\pi\,G\,m_p}$. If we now define the right side of the equation after $f_d$ as 
a function of redshift $g(z)$, we get $\left < \mathrm{DM_{cos}} \right > = f_d\,g(z)$. 
Since we can compute $g(z)$ directly for a given cosmology, we now simply need
to estimate the quantity $\left < \mathrm{DM_{cos}} \right >$ from our data 
in order to produce a measurement $\hat{f_d}$. We go about this in two ways. The first is a simple 
mean, where we include only FRBs beyond redshift 0.2. Each FRB's $f_d$ value is weighted by its inverse variance. We include only higher redshift sources because a larger portion of their total 
extragalactic DM is expected to come from diffuse cosmic gas and $\mathrm{DM_{cos}}/g(z)$ is a more reliable estimator of $f_d$. If we do this, 
we find $\left < \mathrm{DM_{cos}} / g(z) \right > = 995\pm 87$\,pc\,cm$^{-3}$ and 
$\hat{f_d} = 0.93\pm0.08$, where the quoted uncertainty is from propagating 
errors due to subtracting host galaxy DM from $\rm DM_{ex}$ and from the Milky Way contribution.

A shortcoming of this method is its reliance on the $\left < \mathrm{DM}_{\rm host}\right >$ fit from our full
MCMC parameter estimation. If instead we construct a Gaussian likelihood function with,

\begin{equation}
    \mathcal{L} = \prod\limits_i\,\frac{1}{\sqrt{2\pi\sigma_i^2}}\,\exp{\left (- \frac{(\rm DM_{ex,i} - DM^{mod}_{ex, i})^2}{2\sigma_i^2} \right )},
\end{equation}

\noindent we can marginalize over the host DM properties to estimate the 
total diffuse gas content directly. Here, $\mathrm{DM_{ex, i}}$ is the data and $\mathrm{DM^{mod}_{ex, i}}$ is our model for the $i^{th}$ FRB's extragalactic DM, 

\begin{equation}
    \mathrm{DM^{mod}_{ex}} = \mathrm{DM_{cos}}(z_s, f_d) + \left < \mathrm{DM_{host}}\right >(1 + z_s)^{-1}.
\label{eq:likedmex}
\end{equation}

\noindent We estimate several sources of uncertainty that 
contribute to $\sigma_i$ and add them in quadrature. The error on $\sigma_{\rm DM_{ex}}$ is due to subtracting off the Milky Way DM. This is the halo DM uncertainty and the error in NE2001, which we take to be 15\,pc\,cm$^{-3}$ and 
0.1\,$\rm DM_{MW}$ respectively\cite{ocker2020}. The uncertainty in $\mathrm{DM_{cos}}$ is
due to the uncertainty in $\Omega_b\,h_{70}$, assuming that $f_d$ is
the free parameter that we are trying to fit. 
% Thus, $\sigma_{\mathrm{cos}}$
% is only $\approx \,0.01\,\mathrm{DM_{cos}}$ because the baryon content is known at the 1$\%$ level from early Universe measurements\cite{Cooke_2018, Planck2018}. 
The largest source of uncertainty is the host galaxy DM variance, which 
we take to be a free parameter that is marginalized over. 

\begin{equation}
    \sigma^2_i = \sigma^2_{\mathrm{MW}} + \sigma^2_{\mathrm{cos}} + \left ( \sigma_{host}(1+z_s)^{-1} \right )^2 
\end{equation}

\noindent The results of this fit are shown in Extended Data Figure~4. We find \fd for the full sample. For the DSA-110 only sample, it is \fddsaonly. The uncertainty from direct fitting is less than if we take $f_{\mathrm{IGM}} + f_X$ from our previous method, because in that case the error is dominated by model uncertainty in apportioning $\rm DM_{cos}$ to halos vs. intergalactic gas. Our data 
suggest that a large portion of the cosmic baryons are in a diffuse ionized state that can impact FRB DMs.

It is important that these general methods agrees with our primary MCMC fit, because they
do not rely on the relative values of $f_{\rm IGM}$ and $f_X$, 
nor on the statistical scaffolding of IllustrisTNG that was used in our broader inference approach. We are simply estimating the mean cosmic dispersion measure, $\left < \rm DM_{cos} \right >$, normalizing by known a cosmological quantity $g(z)$ to get $f_d$, and then marginalizing over the host contribution. The sizeable $f_d$ found by our methods constrain from above the baryon budget that can be allotted to stars and cold gas.

\noindent{{\bf A late-Universe measurement of $\mathbf{\Omega_b\,h_{70}}$}} If one has an external prior on the diffuse baryon fraction $f_d$, then $\left < \rm DM_{cos} \right >$ can be used to constrain the physical baryon density $\Omega_b\,h_{70}$ independently of early Universe measurements. 
We start by writing down a Gaussian likelihood function on the 
difference between the mean extragalactic DM of our sample and 
the predicted mean $\rm DM_{ex}$ for a given $\Omega_b\,h_{70}$,

\begin{equation}
    P(\left < \mathrm{DM_{ex}} \right >) \propto e^{-\frac{\left ( \left < \mathrm{DM_{mod}} \right > - \left < \mathrm{DM_{ex}} \right > \right)^2}{2\,\sigma^2}}
\end{equation}

\noindent where $\left < \rm DM_{ex} \right >$ comes from our data and,

\begin{equation}
    \left < \mathrm{DM_{mod}} \right > = \left <  f_d\,\Omega_b\,h_{70}\, g(z) + \mathrm{DM_{host}} (1 + z_s)^{-1}\right >.
\end{equation}

\noindent We calculate $\sigma$ based on the 
uncertainty in the mean host contribution 
from our global fit of the cosmic gas parameters, 
uncertainty in $\rm DM_{ex}$, and 
the width of our prior distribution on $f_d$. 
This results in \omegabh.

We can also use the likelihood function from Eq.~\ref{eq:likedmex} and sample 
the posterior via MCMC.
% Assuming that 
% $f_d = 1 - f_* - f_{cold}$, we can construct a prior based 
% on what is known about stellar mass and cold gas. 
% The fraction of baryons in cold gas is about 1.25$\%$ and $\sigma_{f_{cold}} = 0.0025$, i.e. the cold gas uncertainty is negligible (see a later Methods section for more detail). The dominant source of uncertainty in $\sigma_{f_*}$ is not measurement error but model uncertainty in the low-mass behaviour of the IMF. For the sake of conservatism, we assume a flat prior on stellar mass ranging a factor of five, allowing values of $f_* = 0.03-0.15$. This results in a prior on the 
% diffuse fraction of $f_d = \text{Uniform}(0.84, 0.96)$ once accounting for $f_{cold}$. 
The MCMC 
fit estimates the host galaxy DM, its variance, and 
$\Omega_b\,h_{70}$ with the latter posterior 
being $0.049^{+0.04}_{-0.04}$. In Figure~\ref{fig:omega_b_fig} we plot both 
methods. The key difference is that in the MCMC method the likelihood function is over 
all individual FRBs and is free to choose 
a best fit host-DM distribution.
We opt to present the more conservative Gaussian likelihood approach on $\left < 
\rm DM_{ex} \right >$ as our primary 
measurement of the baryon parameter.

\noindent {\bf Partitioning the IGM and Halos} We devise a new method of modelling 
the extragalactic DMs of FRBs. 
Past analyses have used a single 1D PDF to describe the probability distribution of cosmic dispersion, $P(\mathrm{DM}_{cos} | z_s)$, which does not explicitly separate the gas into the 
IGM and halos. 
We choose to describe the cosmic DM distribution as a 2D PDF in $\mathrm{DM_{IGM}}$ and $\mathrm{DM}_{X}$, allowing us to parameterize the fraction of gas in the IGM ($f_{\mathrm{IGM}}$) and halos ($f_X$) in our model. $\mathrm{DM_{IGM}}$ and $\mathrm{DM}_{X}$ are covariant because sightlines that traverse overdensities in the large-scale structure are more likely to intersect halos; conversely, sightlines that pass through voids will have less $\rm DM_{IGM}$ and $\rm DM_X$, on average (see Extended Data Figure~3). Since their PDFs do not factorize, one cannot simply draw from the distributions of halo DM and IGM DM independently. We model $P_{cos}(\mathrm{DM_{IGM}}, \mathrm{DM_{X}} | z_s, \Gamma_{cos})$ as a bivariate
log-normal distribution. This function is log-normal in both $\mathrm{DM_{IGM}}$ and $\mathrm{DM_{X}}$, but with some covariance 
between them. The parameters of the cosmic DM distribution are the log-normal means and standard deviations of the two variables ($\mu_{DM_{IGM}}$, $\sigma_{DM_{IGM}}$, $\mu_{DM_{X}}$, $\sigma_{DM_X}$) and $\rho$, which is the correlation between IGM and the halo contribution. These parameters are all redshift dependent and we calibrate them to IllustrisTNG, as described below. 
Following previous work\cite{macquart20}\cite{jamespdmz}, we model the host galaxy DM in the rest frame as a log-normal distribution, $P_h(\mathrm{DM}_{host} | z_s, \mu_{host}, \sigma_h)$. The mean, median, and variance of the host contribution is $e^{\mu_{host} + \sigma^2_{host}/2}$, $e^{\mu_{host}}$, and $ [e^{\sigma_{host}^{2}}-1]e^{2\mu_{host} +\sigma_{host} ^{2}}$, respectively.

For an input $\mathrm{DM}_{ex}$, host redshift $z_s$, model parameters $\Gamma_{cos}$, $\mu_{host}$ and $\sigma_{host}$, we can compute the likelihood
of a single FRB as

\begin{equation}
    P(\mathrm{DM_{ex}} | z_s, \Gamma) = \int\limits_{0}^{\mathrm{DM_{ex}}}\!\!\!\!\!\!\!\!\!\!\!\!\!\!\!\!\!\!\!\!\int\limits_{0}^{\,\,\,\,\,\,\,\,\,\,\,\,\,\,\,\,\,\,\,\,\,\,\,\,\,\,\,\,\mathrm{DM_{ex}} - \mathrm{DM_{IGM}}} \!\!\!\!\!\!\!\!\!\!\!\!\!\!\!\!\!\!\!
    P_{cos}(\mathrm{DM_{IGM}} , \mathrm{DM_{X}} | z_s, \Gamma_{cos})
    P_h(\mathrm{DM}_{host} | z_s, \mu_{host}, \sigma_h)
    \,\,\,
\mathrm{dDM}_{X}\,\mathrm{dDM}_{\rm IGM},
\label{eq:like}
\end{equation}

\noindent $$\mathrm{DM_{host}}= \left [ \mathrm{DM_{ex}}-\mathrm{DM_{IGM}}-\mathrm{DM}_{X} \right ] (1 + z_s).$$

\noindent We next want to fit our dataset (pairs of $z_s$ and $\mathrm{DM_{ex}}$ values) to physical parameters that describe the distribution of ionized gas in the Universe. Rather than trying to fit the large number of redshift-dependent parameters described previously, we use IllustrisTNG as a baseline\cite{Nelson2019}. Taking results from a mock FRB survey in TNG300-1, we can fit a bivariate log-normal distribution to the
simulated DMs because $\mathrm{DM}_{X}$ and $\mathrm{DM_{IGM}}$ are recorded for each sightline. This was
done at all redshifts, giving us a continuous 2D function $P_{cos}(z_s)$. We found a systematic mis-estimation of DM variance in the ray-tracing methods of Zhang et al. (2020)\cite{ZhangIGM} and Walker et al. (2024)\cite{walker2023}, which is due to redshift gaps between simulation snapshots and the method of interpolation between those snapshots. Previous methods find approximately correct mean DM($z_s$), but must be adjusted for accurate redshift-dependent variance by ray-tracing without gaps between snapshots (Konietzka et al., in prep, will explore this further). 

We then vary the $P_{cos}$ distribution from TNG300-1 by making
$f_{\mathrm{IGM}}$ and $f_X$ free parameters that can be fit to our data. Halos are defined as regions where the dark matter density is at least 57 times the critical 
density of the Universe\cite{martizzi2019, artale2022, walker2023}. This is the expected dark matter overdensity at $R_{200}$. Since we know the TNG300 values\cite{walker2023} of $f_{\mathrm{IGM}}$ and $f_X$ as well as $\mu_{\rm DM_{IGM}}$ $\mu_{\rm DM_{X}}$, we can
calibrate to the simulation without being restricted by the particular partition in Illustris. We transform $\mu_{\rm DM_{IGM}}$ and $\mu_{\rm DM_{X}}$ in the following way,

\begin{align*}
    \mu_{\rm DM_{IGM}} &= \mu_{\rm \mathrm{DM_{IGM, TNG}}} + \log{\frac{f_{\mathrm{IGM}}}{0.827}} \\
    \mu_{\rm DM_{X}} &= \mu_{\rm \mathrm{DM_{X, TNG}}} + \log{\frac{f_{\mathrm{X}}}{0.138}}
\end{align*}

\noindent For example, if one increases $f_{\rm X}$ by $10\%$,
the mean DM from intervening halos will increase by $10\%$. 
While the log-variance is fixed at the value 
of our baseline simulation, the variance is not. Increasing 
$f_{\rm X}$ will also increase the variance from halos, because the variance of a log-normal distribution depends on both its log-mean and log-variance ($\left [ e^{\sigma_{DM_{IGM}}} - 1 \right ] e^{2\mu_{\rm DM_{IGM}} + \sigma_{DM_{IGM}}^2}$). As expected, if halos are more gas-rich, the total DM and line-of-sight scatter from halos will increase.

By using TNG300-1 as a baseline and allowing $f_{\rm IGM}$  and $f_{X}$ to be free parameters, we are effectively adopting the definition of the IGM and halos used in the mock FRB survey that we calibrate against\cite{walker2023}. They follow previous attempts at partitioning the large-scale structure\cite{martizzi2019,artale2022} and assume three classes of cosmic structure: Voids, with $\rho_{dm}/\rho_c < 0.1 $; filaments with $0.1 < \rho_{dm}/\rho_c < 57$; and halos with $\rho_{dm}/\rho_c > 57$. We combine filaments and voids to be the IGM. Our results therefore assume $\rho_{halos} > 57\rho_c$ and $\rho_{\rm IGM} < 57\rho_c$. 

Several of the 2D $P_{cos}$ distributions with different parameters are shown in the top two rows of Extended Data Figure~3 for $z_s=0.5$ and $z_s=1$. 
The bottom row shows the resulting Macquart relation 
of that columns' cosmic gas parameters.
As gas moves from the IGM to halos (left to right columns), the total line-of-sight DM variance increases.

We seek to compute a posterior distribution over four parameters $\mathbf{\theta} = \left \{ f_{\mathrm{IGM}}, f_{X}, \mu_h, \sigma_h \right \}$. The posterior is proportional to the likelihood from Eq.~\ref{eq:like} multiplied by our prior distribution on $\vec{\theta}$,

\begin{equation}
    P(\, \theta \, | \, \mathrm{DM}_{ex}, z_s) \propto P(\, \mathrm{DM}_{ex}, z_s\, | \, \theta \, )\times P(\theta).
\end{equation}

\noindent Taking its logarithm and ignoring the constant offset from the evidence term, we get

\begin{equation}
    \log P(\, \theta \, | \, \mathrm{DM}_{ex}, z_s) = \mathcal{L} + \log P(\theta)
\end{equation}

\noindent where $\mathcal{L}$ is the total log-likelihood that comes from 
summing over all $n_{FRB}$ sources,

\begin{equation}
    \mathcal{L} = \sum_{i}^{n_{FRB}}\,\log P(\mathrm{DM}_{ex, i} |  z_{s,i} \, \theta).
\end{equation}

\noindent We are effectively treating each FRB's $(\mathrm{DM}_{ex}, z_{s})$ pair as independent by summing their 1D log likelihoods. While a likelihood function need not necessarily sum to one, as they are not PDFs, we normalize $P(\mathrm{DM}_{ex} | z_{s})$ to have the 
same integral at each $z_s$. An alternative approach would be to compute a single 2D likelihood function that takes into account the FRB redshift distribution and various observational selection effects\cite{jamespdmz}---in other words, the full forward model. We choose not to do so because the completeness of DSA-110 as a function of DM, pulse width, scattering, and fluence has not yet been fully characterized. An exception is that each $P(\mathrm{DM}_{ex, i} |  z_{s,i})$ for DSA-110 FRBs has been multiplied by a Heaviside function that is zero above the maximum search DM during commissioning of 1500\,pc\,cm$^{-3}$. Below that value, we recover injected simulated FRBs at consistent rates. We estimate that this is only 
marginally sub-optimal compared to weighting by the full DM sensitivity curve\cite{connor2019, jamespdmz}.

\noindent {\bf The DM Cliff} We find evidence of a ``DM cliff'', first described by\cite{jamesH02022}, which is the sharp 
probability cutoff at low $\rm DM_{ex}$ for a given $z_s$. 
It is much more common for an FRB to be over-dispersed (i.e. above the maximum-likelihood 
DM for a given redshift) than under-dispersed. This is because 
the FRB can intersect a galaxy cluster, traverse the barrel of a filament, or be significantly locally dispersed, but if the sightline 
passes through one or multiple voids and does not intersect any halo, 
the IGM will still contribute a baseline quantity of cosmic DM\cite{walker2023}. 
Said more plainly, the distribution $P(\mathrm{DM_{ex}} | z_s)$ is skew 
with a wide tail at high DM$_{ex}$ values. 

Sources near the low-end in DM, or the ``DM Cliff'', place the tightest constraints on the IGM. For FRBs in our sample beyond redshift 0.1, we do not find any source with $\mathrm{DM_{ex}} / z_s < 800 $\,pc\,cm$^{-3}$. If the IGM were depleted and most 
extragalactic DM came from the intersection of halos, we would not expect such 
a sharp fall-off at low DM$_{ex}$ at a given redshift. An FRB at $z_s\approx0.5$ can intersect zero, one, or a few halos, which would lead to large Poissonian variance and this cliff would be both lower and smoother. The same is true if 
a dominant portion of the dispersion budget were due to a wide
distribution of local DM, as this would smear out $P(\mathrm{DM_{ex}} | z_s)$.
The errors we quote on the inferred DM parameters are the $1\,\sigma$
width of our MCMC posteriors times 1.3. We add 30$\%$ to the uncertainty 
as a model uncertainty term because we have only used one simulation as a 
baseline, leading to systematic error. The value was motivated 
by the 30$\%$ increased uncertainty in the fit parameters when we 
force a model mismatch between the assumed TNG parameters and the true values.

\noindent {\bf Verification of DM/$z_s$ method} Given the novelty of our approach to fitting extragalactic DM, we wish to further verify the method and the results it has produced. We have already established that a partition-independent 
analysis of mean cosmic DM reproduces a large $f_d$; we would also like to know if the large $f_{\rm IGM}$ 
value is supported by other methods.

We start with a simple argument 
against a low value of $f_{\rm IGM}$ and high halo gas content, $f_X$. Suppose the IGM 
were maximally smooth (i.e. a one-to-one mapping between 
redshift and DM$_{\rm IGM}$). If $f_{\rm IGM}$ were 
0.50, the lowest possible extragalactic DM would be 
roughly 1080\,$f_{\rm IGM}\,z_s$\,pc\,cm$^{-3}$ = 540\,$z_s$\,pc\,cm$^{-3}$, corresponding to sightlines 
that do not pass through a halo. The majority of our sample does not pass through a halo, as $\tau_{halo}\approx1$ at $z_s=0.6$ and the median redshift of our sample 
is 0.28. If $f_{\rm IGM}$ were 0.50
and the cosmic baryons traced dark matter (i.e., maximally porous on large scales with a dark matter/gas bias of one),
then the Macquart Relation would have a large scatter and many sightlines would have $\rm DM_{ex} < 540\,z_s$. The two extremal low $f_{\rm IGM}$ scenarios are ruled out by our data, where $\mathrm{DM}_{ex} > 800\,z_s$ for all sources beyond redshift 0.1.

In future work, a different cosmological simulation (e.g. SIMBA or EAGLE) could be used as a baseline model from which to do simulation based inference. As an approximation to that, we can alter the DMs in the TNG300-1 
mock FRB survey to artificially increase the effective $f_{X}$ and decrease $f_{\rm IGM}$. Each sightline has some DM$_X$, DM$_{\rm IGM}$, and redshift. We simply multiply 
all DM$_X$ by a fixed value and decrease all DM$_{\rm IGM}$ 
values such that $f_{X}$ + $f_{\rm IGM}$ is conserved. 
This is akin to moving baryons into halos and out of the 
cosmic web, but using the same $f_d$ and $\Omega_b\,h_{70}$.
An example is shown in panel ``\textbf{a}'' of Figure~\ref{fig:pdmsims}, where we compare the 
DM distributions for simulated FRB sources at $z_s=1$ using 
two different sets of cosmic gas parameters. Although the 
two distributions have the same mean DM, there is a clear 
difference at DMs below $\approx1000$\,pc\,cm$^{-3}$ (in other words, the 
location and shape of the DM cliff is highly sensitive to $f_{X}/f_{\rm IGM}$).

In panel ``\textbf{b}'' of Figure~\ref{fig:pdmsims}, we compare our localized FRB sample with the two simulated scenarios, 
$(f_{X}=0.32, f_{\rm IGM}=0.60)$ and $(f_{X}=0.12, f_{\rm IGM}=0.80)$, and 
one semi-analytic case where the baryon and dark matter distributions are identical. We have taken the modified IllustrisTNG 
DMs for all sources $0.20<z_s\leq1.5$, added to them a log-normal host DM distribution with 
$\mu_{host}=4.8$ and $\sigma_{host}=0.5$ weighted 
by $(1+z_s)^{-1}$, and then excluded high-DM sources that 
would not have been detected by our instruments. This is a 
crude forward model for the observed DM distributions 
in a baryon-rich IGM scenario ($f_{\rm IGM}=0.80$) vs. a baryon-deficient cosmic web ($f_{\rm IGM}=0.60$). Next, we normalize 
these DM$_{ex}$ values by $g(z)$ (defined after Equation~\ref{eq:dmcos}) and multiply by 1080\,pc\,cm$^{-3}$, which is the mean cosmic DM at $z=1$ if $f_d=1$. This allows us to compare across multiple redshifts. 
For the case where baryons trace dark matter, 
we have taken $\sigma_{DM}(z_s)$ from Mcquinn (2013)\cite{mcquinn13}, which was 
calculated analytically by integrating the dark matter powerspectrum.
We combine that with $\mathrm{DM}(z_s)$ assuming baryons trace dark matter. We then plug those values 
into the functional form for $P(\mathrm{DM} | z_s)$ used by Macquart et al. (2020)\cite{macquart20} and assume the same host galaxy DM distribution 
as before to estimate $P(<\rm DM)$ for sources between $0.20<z_s\leq1.5$.
This gives the dotted blue curve shown in panel ``\textbf{b}'' of Figure~\ref{fig:pdmsims}.

We plot our 
FRB sample in black for all sources beyond redshift 0.20. 
We compute a two-sample Kolmogorov–Smirnov (KS) test 
comparing our observations with the simulated FRB DMs. 
We find $p_{KS}=9\times10^{-4}$ when comparing our data with 
the $f_{\rm IGM}=0.60$ sample and $p_{KS}=0.09$ for the $f_{\rm IGM}=0.80$ data. The point here is not
to obtain a fit of the cosmic gas parameters in this space, but instead to show that our central results can be produced 
independently of the multivariate log-normal distribution 
and the MCMC posterior estimation. The case where the IGM is devoid of baryons and the halos have roughly the cosmological average of $\frac{\Omega_b}{\Omega_m}\approx0.16$ is disfavoured out by our data. 

There are risks associated with simulation-based inference, such as overfitting to the particular simulation that was used or subtleties in ray-tracing. As discussed, we seek to extend this approach to different cosmological simulations in the future as well as suites with a variety of feedback and cosmological parameters. It will also be crucial to develop analytic tools, even if they cannot capture the full complexities of astrophysical feedback and the large-scale baryon distribution. A basic halo-model approach to 
DM statistics may still prove valuable. 

\noindent {\bf The host DM distribution}
We have paramaterized the host DM distribution as log-normal 
with $\log{\mathrm{DM_{host}}}\sim\mathcal{N}(\mu_{host}, \sigma_{host}^2)$. 
The median host DM of such a distribution will be $e^{\mu_{host}}$, which we 
find to be modest for our cosmological sample, at \dmhost\,pc\,cm$^{-3}$. The mean $\mathrm{DM}_{host}$ of our best fit is $\sim$\,30$\%$ higher than the median. This is in line with other analyses\cite{Shin2023, flimflamdr1} that 
find the median rest-frame host DM of detected FRBs is likely not much more than $\mathcal{O}(10^{2})$\,pc\,cm$^{-3}$.
If one considers the most nearby FRBs (several of 
which did not meet our sample criteria), it is not uncommon to have low 
host DM. FRB\,20200120E resides in an M81 globular cluster with 
almost no $\rm DM_{host}$ beyond the M81 CGM\cite{M81mohit,M81Kirst}; FRB\,20220319D is in a spiral 
galaxy at just 50\,Mpc and likely has less than 10\,pc\,cm$^{-3}$ from the host\cite{ravi2022mark}; FRB20181030A is at $z_s=0.0039$ and likely has a local contribution below 30\,pc\,cm$^{-3}$\,\,\cite{Bhardwaj2021b}; and the periodically active repeating source 
FRB\,20180916B at $\sim$\,150\,Mpc has no detectable local scattering even down 
to 100\,MHz\cite{R3marcote, R3_scattering_pastor}, indicating a relatively pristine nearby environment. 
The modest host DMs found in our and others' works may 
suggest older stellar populations as the origin of many FRBs\cite{orr2024, kovacs2024dm}. 
There are of course counter examples (FRB\,20121102A, FRB\,20190520B, FRB\,\frbada, etc.), but we now have evidence that many observed FRBs reside in typical
locations in their galaxies, away from significant star formation\cite{SharmaNature2024}. Since we include the 
host halo's contribution in $\mathrm{DM}_{host}$, 
the local Universe sample supports our broader finding 
that the CGM of $L^*$ galaxies are not baryon rich. 

We naively expected most of the constraining power in our 
four-parameter MCMC fit of the baryon distribution to come 
from the most distant sources. This may be true for $f_{\rm IGM}$ and $f_X$, 
but more nearby sources ($z_s \lesssim 0.2$) provide a useful
anchor for the host DM distribution. The fit ``learns'' the host 
DM distribution from low redshift sources because it is a larger fraction of the total DM and because of the $(1+z_s)$ dilution factor of the restframe DM in the host galaxy. Indeed, 
when we include only sources beyond redshift 0.2, 
the uncertainty on $\mu_{host}$ and $\sigma_{host}$ 
increases dramatically with best fit values of
$3.0_{-2.0}^{+1.4}$ and 
$1.1_{-0.6}^{+0.8}$, respectively. The cosmic gas parameters 
remain consistent at the 1\,$\sigma$ level with the main sample, but $f_{IGM}$ increases by about $4\%$.

Our primary analysis method does not include redshift dependence in the host 
contribution. This is because the median redshift of FRBs with 
host galaxies is $\sim\,0.28$, with 90$\%$ of sources at $z_s\leq0.6$. Thus, we 
assume a single log-Normal distribution for the rest-frame DM. However, 
for a larger sample with many sources beyond redshift 1, the evolution of 
host galaxies and their ISM may become important. 

\noindent {\bf Resampling and jackknife tests} Our FRB sample spans a factor of roughly 200 in luminosity distance, with sources from $\sim$\,50\,Mpc away and several sources at 
or beyond redshift 1. We have FRBs detected and localized by different interferometers, as well. It is therefore worth testing how our parameter estimation responds to different 
subsets of the data. We have carried out several 
resampling tests, starting with running MCMC fits with different redshift cuts. If we exclude from our standard sample the high-redshift sources ($z_s > 0.8$), we find 
\figmlowz, \fXlowz, \muhostlowz, and \sigmahostlowz. When analyzing only the DSA-110 detected sample, we get \figmDSA, \fXDSA, \muhostDSA, and \sigmahostDSA. All are consistent at the 1-$\sigma$ level with our primary sample fits.

We have tried reintroducing FRB\,20190520B and FRB\,\frbada to our fit, which were not used in our primary sample because of their inferred high local DM. Their inclusion does not significantly alter 
the cosmic gas parameters $f_{\rm IGM}$ or $f_X$, which are 
$0.81_{-0.10}^{+0.09}$ and $0.21_{-0.10}^{+0.11}$ respectively.
The central value of $f_X$ increased, but was still consistent 
with other sample results within the uncertainty. The uncertainty on $f_{\rm IGM}$ was marginally higher.
As expected, inclusion of the local-DM sources increase both 
the mean host DM and $\sigma_{host}$ slightly. The uncertainty on $f_X$, $\mu_{host}$, and $\sigma_{host}$ all 
increased. This is a useful sanity check, establishing that
our model behaves as expected and can differentiate 
between the IGM and other components.

Jackknife resampling for error 
and bias estimation involves excluding one observation from the sample per fit and estimating parameters for each subset. This typically requires independent and identically distributed ($i.i.d$) samples. In our case, 
each FRB does not carry equal information: Higher redshift FRBs carry more information about the cosmic gas than low redshifts sources and FRBs near the 
``DM Cliff'' will do more work constraining $f_{\rm IGM}$. With those caveats aside, we have removed each source, run the MCMC fits, and collected the statistics of the posterior estimates less one FRB. The jackknife uncertainty on parameter $\theta$ is $\sigma_{\theta} = \sqrt{n_{FRB} - 1}\,\,\sigma_{JK}$, where $\sigma_{JK}$ is the standard 
deviation of the jackknife results. We find 
$\sigma_{f_{\rm IGM}}=0.09$, slightly larger than the error on our full-sample fits. We obtain similar results for the other parameters. We choose to present our full-sample MCMC posteriors as our primary results because our data 
do not match the standard criteria for jackknife error estimation.

In addition to resampling tests, we can change our prior on $f_{\rm IGM}$. We initially wanted to keep our priors as wide as possible and allowed the $f_{IGM} + f_X$ to exceed 1 slightly for flexibility (if there were issues with the model, the data may end up preferring unphysically large values, etc.). If instead we set an upper-limit on the total cosmic 
gas based on the measurement of $f_d$, we can use $f_{IGM} + f_X \leq 0.98$. The lower ceiling on $f_{IGM} + f_X$ leads to smaller error bars on the high side of the $f_{IGM}$, but a similar central value.

\noindent {\bf Comparison with past methods} Ever since the discovery of FRBs, it has been a central goal to use their DMs to map out the Universe's baryons, which was made possible with the first sample of sources localized to their host galaxy\cite{macquart20, yang22}. Past methods have modelled the total cosmic dispersion (DM$_{IGM}$ + DM$_{X}$), because of limited sample sizes and the lack of analysis tools for partitioning the cosmic gas. This has allowed Macquart et al. (2020)\cite{macquart20} and Yang et al. (2022)\cite{yang22} to constrain the total diffuse baryon content of the Universe. The latter work include multiwavelength observations assembeled by Shull et al. (2012)\cite{shull2012}, but do not partition the cosmic gas with FRB data. 
Previously, it was assumed that the majority of
line-of-sight variance in FRB DMs comes from intersecting foreground galaxy halos\cite{mcquinn13}\cite{macquart20}, 
which can be described as a Poisson process. This would lead 
to the natural expectation that the fractional standard deviation of DM, $\sigma_{DM}$, scales as $z_s^{-1/2}$ because 
the number of halos intersected is roughly proportional to redshift. 
If this redshift scaling is known, 
then the line-of-sight variance can be parameterized by 
a single number. Indeed, a 
parameter $F$ (a ``feedback'' or ``fluctuation'' parameter) has been proposed\cite{macquart20}\cite{jamespdmz, baptista23}, which is defined as $F\equiv \sigma_{DM}\sqrt{z_s}$ and is approximated as a fixed, 
redshift-independent value. A virtue of the $F$ parameter is its simplicity: a Universe with a large fluctuation $F$ will have large line-of-sight variance in DM, caused by weak feedback\cite{baptista23}; small $F$ results in a smooth Universe and low $\sigma_{DM}$. In this scenario, halos retain their baryons and the Universe is less homogeneous. 

Several works have adopted the following 
analytic form for cosmic $\rm DM_{cos}$\cite{macquart20}\cite{ZhangIGM},

\begin{equation}
    P_{cos}(\Delta) = A\,\Delta^{-\beta} \exp{\left [ -\frac{(\Delta^{-\alpha} - C_0)^2}{2\alpha^2\,\sigma^2_{DM}} \right ]},
\end{equation}

\noindent which includes both halos and the IGM. Here, $\Delta(z) \equiv \frac{\rm DM_{cos}}{\left < \rm DM_{cos }\right >} $ and the effective standard deviation is 
given by $\sigma_{DM} = F\,z^{-1/2}$. Most authors then assume 
$\alpha=3$ and $\beta=3$ based on the premise that 
the variance in $\Delta$ comes from halos and that the gas in halos have some assumed radial distribution. 

While this is a useful parameterization of the total gas distribution, 
the premise that DM variance 
is due mainly to the Poissonian intersection of intervening halos may not hold, requiring a redshift-dependent $F$ parameter. We find that significant scatter in 
extragalactic DMs comes from the large-scale structure in the IGM, particularly the intersection of filaments and sheets, which can occur at a variety of angles and is less well described by Poisson statistics. The intersection of filaments, sheets, and voids is also less classically Poissonian because of the large number of intersections per sightline ($\mathcal{O}(10^2)$). This is 
borne out in simulations such as IllustrisTNG, where it has been observed that 
$\sigma_{DM}$ does not scale as $1/\sqrt{z_s}$~\cite{baptista23}, indicating that the DM scatter is not a simple Poisson process.
Our simulations found that the intersection of filaments plays the dominant role in determining cosmic DM. This is also explicitly demonstrated in mock FRB surveys\cite{walker2023} where 
the IGM contributes significantly more DM$_{ex}$ than halos and considerable line-of-sight variance. The Macquart Relations for each component of DM is shown in Extended Data Figure~2. 

We are interested in using FRB DMs to partition the baryons into the IGM and halos. For this reason, we require a
fitting method that explicitly parameterizes the location of cosmic gas. The de facto standard method for modelling  DM/$z_s$ with $P(\Delta)$ does not provide a partition. Stacking analyses have been used to search for the impact of foreground halo gas on unlocalized CHIME/FRB sources\cite{connorravi22}$^{,}$\cite{wumcquinn23}, but these do not measure the IGM content. Another method that has emerged for interpreting the extragalactic DMs of FRBs is centered around foreground mapping\cite{kglee22}. Wide-field galaxy survey data are used to reconstruct intervening large-scale structure, with the goal of using this model to estimate the baryon partition on a per-source basis. A recent study of eight localized FRBs by the FLIMFLAM collaboration employed the foreground mapping method, finding a significantly
lower value of $f_{\rm IGM}$ and lower $f_d$ than our results\cite{flimflamdr1}. 
% Part of this discrepancy 
% may come down to definitions of the IGM and assumptions about the cold baryon fraction. For example, if their effective definition of halos is gas 
% in regions where the dark matter density
% is greater than $10\,\rho_c$ ($R \lesssim 3.1\,R_{vir}$ for NFW), then this would decrease $f_{\rm IGM}$ considerably. Still, 
It will be useful to 
compare global fitting methods such as ours with foreground mapping 
efforts of larger numbers of sightlines.

In our and others' scheme\cite{artale2022, martizzi2019, walker2023} the 
reference density is $57\,\rho_c$.
We find that using a reference density closer 
to the virial density is appropriate, because otherwise dense
plasma in filaments and sheets, far from collapsed halos, would incorrectly contribute to $f_{X}$, not $f_{\rm IGM}$ (see Figure 1 in Walker et al. (2023)\cite{walker2023}). Ultimately, one wants to characterize the baryon powerspectrum\cite{Madhavacheril2019, Reischke24} in order to measure gas fluctuations on all scales, without requiring any taxonomy.

\noindent {\bf Baryon fraction in halos} The dark matter halos of individual galaxies are filled 
with multi-phase gas known as the circumgalactic medium (CGM)\cite{TumlinsonCGM}.
The shared gas in the halos of galaxy groups
is the intragroup medium (IGrM). The ionized plasma in the 
most massive halos is the intracluster medium (ICM) 
and is the best observationally constrained of the three, 
thanks to the detectability of the ICM in X-ray ($\propto \!\int n_e^2\,dl$) and 
SZ ($\propto\!\int n_e\,T_e\,dl$)\cite{eRASS1-clusters2024, psz2, act_clusters}. While the taxonomy varies throughout the literature,
we define the ICM as ionized gas in halos 
with $M \geq 10^{14}\,h_{70}^{-1}\,M_\odot$,
the IGrM as gas within halos $10^{12.7}\,h^{-1}_{70}\,M_\odot \leq M \leq10^{14}\,h^{-1}_{70}\,M_\odot$, 
and the CGM as gas in halos with $M\leq10^{12.7}\,h^{-1}_{70}\,M_\odot$. Technically, 
the halos of modest galaxy groups such as the Local Group would be classified as CGM in this definition\cite{LocalGroupMass}.

Approximately $14\%$ of the total matter in the 
low-redshift Universe resides within $r_{200}$ of halos with mass greater than 
$10^{13}\,h^{-1}_{70}\,M_\odot$\cite{bohringer2017}. 
For galaxy clusters with $M\geq10^{14}\,h^{-1}_{70}\,M_\odot$, 
this number is $4.4\pm0.4\,\%$. The baryon gas fraction in halos,
$f_{gas}$, is a function of halo mass, and approaches the cosmic 
value of $\sim$\,0.16 for the most massive clusters. 
This quantity is well measured by X-ray and SZ observations for galaxy clusters\cite{Sun2009}, 
but less well constrained for groups. It is even more difficult to pin down for the gas fraction of individual galaxies. Below
we amalgamate multiple public datasets to estimate the fraction of
cosmic baryons in halos of different masses. 
We estimate the fraction of baryons in halos above $M_h$ by integrating the cluster mass function, $n_{cl}(M)$, weighted by the mean hot baryonic fraction in those halos $f_{hot, >M_h} = \frac{1}{\rho_{c}\,\Omega_b}\int^{\infty}_{M_{h}}\,n_{cl}(M)\,f_{hot}(M)\,M\,\mathrm{d}M$. Both $f_{hot}$ and $n_{cl}(M)$ 
are now known more precisely than ever.

\noindent {\it Clusters}: The total baryonic material contained in the ICM
can be estimated with the cluster mass function, $n_{cl}(M)$, 
and the mean baryonic fraction in those halos $f_{gas}(M)$. 

\begin{equation}
    f_{\mathrm{ICM}} = \int^{\infty}_{M_{cl}}\,n_{cl}(M)\,f_{gas}(M)\,\mathrm{d}M
\end{equation}

In Fukugita, Hogan, \& Peebles (1998)\cite{fukugita1998}, the authors took the 
cluster mass function from\cite{bahcallcen93} and 
estimated $\Omega_{\mathrm{HII}, cl} = 1.55^{+1.0}_{-0.72}\times10^{-3}$\,$h^{-1.5}$, 
which translates to $f_{\rm ICM}\approx1.7-5.2\,\%$ with 
modern values of $h$ and $\Omega_b$. Based on an 
updated definition of cluster mass, it was  
estimated\cite{fukugita2004} that $4\pm1.5\,\%$ of the Universe's baryons reside 
in galaxy clusters. This value was adopted in an census of cosmic baryons from just over one decade ago\cite{shull2012}.

In the decades since the first estimates of $\Omega_{\rm ICM}$, 
samples of galaxy clusters have grown considerably\cite{eRASS1-clusters2024}.
Measurements of $f_{gas}(M)$ have improved in precision and have also crept down to lower and lower halo masses\cite{Sun2009, DaiBregman2009}. The REFLEX galaxy cluster survey, comprised of 911 X-ray luminous ROSAT
clusters, was used to determine the mass function of 
galaxy clusters to $\sim$\,10$\%$ precision. 
They parameterized the cumulative total mass function as,

\begin{equation}
    n_{cl}(>M) = \alpha \left ( \frac{M}{2\times10^{14}\,h^{-1}_{70} M_{\odot}} \right )^{-\beta} \exp{(-M^{\delta} / \gamma)},
    \label{eq:nofcl}
\end{equation}

\noindent with best-fit values of  
$\beta=0.907$, $\gamma=0.961$, $\delta=0.625$, 
and $\alpha=1.237\times10^{-5}\,\mathrm{Mpc}^{-3}\,h^{3}_{70}\,10^{14}\,M_{\odot}$.

Recently, The German eROSITA Consortium (eROSITA-DE)
made public its first data release of the Western Galactic Sky 
in the 0.2—10 keV energy band. The survey produced a galaxy cluster catalog with 12,247 optically confirmed galaxy groups and clusters detected in X-rays\cite{eRASS1-clusters2024}. Their inferred cluster mass function now agrees with 
other cosmological measurements of $S_8$.

Integrating the cluster mass function above $10^{14}\,h^{-1}_{70}$,
we find that $\Omega_{cl} = 4.4\,\pm 0.4\,\%$ of the Universe's matter is confined within $r_{200}$ of massive galaxy clusters. Assuming $f_{b, cl} = 13.5\pm1$, we find that 
$f_{ICM} = 3.75\,\pm 0.5\,\%$ of the cosmic baryons reside 
in galaxy cluster gas. %https://www.aanda.org/articles/aa/pdf/2017/12/aa31205-17.pdf

\noindent {\it Massive groups}: Using the same halo mass function from Eq.~\ref{eq:nofcl}, roughly $10\pm1\,\%$ of 
the total matter in the Universe is in halos between $10^{13}\,h_{70}^{-1}\,M_\odot$ and 
$10^{14}\,h_{70}^{-1}\,M_{\odot}$ 
and $13.5\pm1\,\%$ is between $10^{12.7}\,h_{70}^{-1}\,M_\odot$ and 
$10^{14}\,h_{70}^{-1}\,M_{\odot}$. However, the average baryon fraction of massive galaxy groups is less well constrained than that of 
clusters. Still, there is good evidence that the baryon fraction of halos below $10^{14}\,h_{70}^{-1}\,M_\odot$ falls below the 
cosmic average. We use the following empirical $f_{gas}$/$M_{500}$ relation\cite{Sun2009} for the weighted integral of halo masses,

\begin{equation}
    f_{gas}(M_{500}) = (0.0616 \pm 0.0060) \left ( \frac{M_{500}}{10^{13}\,M_\odot} \right )^{0.135\pm0.030}.
\end{equation}

\noindent Here $f_{gas}$ is the average baryon fraction within $r_{500}$. Integrating the cluster mass function 
with $f_{gas}$, then adjusting from $r_{500}$, we 
find that $5.4^{+1.0}_{-1.0}\%$ of baryons are in the IGrM.

\noindent {\it The circumgalactic medium}: The 10$^{4-7}$\,K gas in the 
halos of galaxies ($0.1<\frac{R}{R_{200}}<1$) plays a significant role in galaxy formation and evolution, but there remains heated debate over its total mass and spatial distribution\cite{TumlinsonCGM}. The low density and high ionization fraction render the CGM difficult to observe directly, but also difficult to model because of the complex astrophysics involved: AGN feedback, stellar feedback, and gravitational accretion shocks, all likely play a role in the distribution of this gas. In our work, we define the CGM as gas around $M\leq 5\times10^{12}\,M_\odot$ halos where the dark matter over-density is at least 57 times the critical density, not including the disk.

Our analysis of FRB DMs combined with the baryon counts of massive halos suggest that the mass of the CGM 
around $L*$ galaxies cannot be a major component of the total cosmic baryon budget. In other words, it is not the case that most of the ``missing baryons'' are hiding in 
galaxy halos. This agrees with a recent stacking analysis of $\mathcal{O}(10^5)$ galaxies in X-ray from the eRASS all-sky survey\cite{cgmerosita2024},
which detected the hot CGM but at a level that indicates a baryon deficit. 
Our measurement of $f_X$ from FRBs cannot differentiate between the CGM and gas in more massive halos. However, if we combine our measurement of $f_{\rm IGM}$ with external measurements of groups, clusters, and cold gas, we can constrain the allowed budget of the ionized CGM. Our findings of a baryon-rich IGM also agree with a cross-correlation between kinematic Sunyaev-Zel’dovich (kSZ) effect
from the Atacama Cosmology Telescope the luminous red galaxy (LRG) sample of the Dark Energy Spectroscopic Instrument (DESI) imaging survey\cite{ksz2024}. The authors claim a $40\sigma$ discrepancy between their data and the scenario where baryons tracing the dark matter at small scales, indicating strong feedback.

The CGM is a complex, multi-phase medium that cannot 
be studied with a single empirical probe.
The best measured phase is the cool CGM thanks to the available UV lines 
at $10^{4-5}$\,K\cite{prochaska2011b, TumlinsonCGMb}.
We estimate the budget of the cool CGM ($\sim$\,10$^4$\,K) based on results from the COS-Halos Survey\cite{werk2014}, which indicate that $2-4\,\%$ of the
baryons in the Universe are in this phase in halos between 
$10^{11}$\,M$_\odot$ and $2\times10^{12}$\,M$_\odot$. In order to 
estimate the total CGM portion of the baryon budget, we combine multiple 
measurements. Noting that $f_{\rm CGM} = f_d - f_{\rm IGM} - f_{\rm ICM} - f_{\rm IGrM}$, the probability distribution of the CGM baryon fraction is,

\begin{equation}
    p(f_{\mathrm{CGM}}) = \int\limits_{0}^{1-f_{cold}} p(f_d)\!\!\!\! \int\limits_{0}^{f_d - f_{\rm CGM}} \!\!\!\! p(f_{\rm IGM}) \!\!\!\! \!\!\!\! \int\limits_{0}^{f_d - f_{\rm CGM} - f_{\rm IGM}} \!\!\!\! \!\!\!\! p(f_{\rm ICM})\,p(f_{\rm IGrM})\,\,
    \mathrm{d}f_{d}\,\mathrm{d}f_{\mathrm{IGM}}\,\mathrm{d}f_{\mathrm{ICM}}.
\end{equation}

\noindent We integrate $f_d$ from 0 to 0.96, which is the largest possible value of $f_d$ given what we know about the cold gas and stars. We use this probability density function to estimate 
$f_{\rm CGM}=0.08_{-0.06}^{+0.07}$. This means that 
on average, \fgas for halos between $10^9\,M_\odot$ and $5\times10^{12}\,M_\odot$.

\noindent {\bf The baryon content of galaxies} The total baryon content of galaxies (stars, stellar remnants, and cold gas) is probed in a multitude of ways\cite{baryoncyclereview2020}\cite{walter20}. The primary uncertainty in $f_{gal}$ is the total mass in stars, $f_*$\cite{conroy2013}. 
Specifically, the choice of IMF dictates the abundance of low-mass stars ($M\leq0.4\,M_\odot$) that make up a large fraction of total stellar mass despite contributing only $1\%$ of the bolometric stellar luminosity\cite{conroy-lowmass2012}. 

\noindent \textit{Cold gas:} The majority of cold gas in the Universe is 
neutral atomic Hydrogen, with traces of molecular Hydrogen and Helium. The fraction of baryons in cold gas is therefore
roughly,

\begin{equation}
    f_{cold} \approx \frac{\rm \Omega_{HI} + \Omega_{H_2} +\Omega_{HeI}}{\Omega_b}
\end{equation}

Each quantity is estimated by a variety of observational means. For example, 
at low redshifts, 21\,cm galaxy surveys measure the HI mass function which 
can then be integrated to estimate the neutral hydrogen density at $z<1$\cite{Zwaan2005}.
%(e.g., Zwaan et al. 2005; Braun 2012; Jones et al.2018)}. 
A very different approach recently made the first detection of cosmological HI via 
intensity mapping\cite{Paul23}, which is in principle sensitive to all neutral Hydrogen and not just that confined to galaxies. Combining their 
HI powerspectrum measurement with priors from other probes, the authors found that $\log \Omega_{\rm HI} = -3.23^{+0.15}_{-0.16}$\cite{Paul23}. 

We take the values for $\rm \rho_{HI}$ and $\rm \rho_{H_2}$ in galaxies, and their associated uncertainty from\cite{walter20}, who consolidated a large suite of volumetric surveys to study galaxy-associated gas over cosmic time. Converting to cosmological density parameters we find $\Omega_{\rm HI} = 4.7^{+1.9}_{-1.1}\times10^{-4}$ and $\Omega_{\rm H_2} = 7.9^{+4}_{-2}\times10^{-5}$ at $z=0$. This suggests 
that roughly one percent of the baryons are 
in cold gas within galaxies. 

\noindent \textit{Stars:} The $\rho_*$ and $\Psi_*$ values from\cite{madaudickinson2014} use a Salpeter IMF, which has significant probability weight at low stellar masses. They integrate between 0.1 and 100\,$M_\odot$. In \cite{walter20} the authors fit a smoothly varying function 
to the same noisy data as a function of redshift, finding that 
$\rho_* = 4_{-0.8}^{+1.7}\times10^{8}\,(\mathrm{M}_\odot\,\mathrm{Mpc}^{-3})$ on the fitted curve at $z\approx0$ (see Figure 2. of that paper). As a baryon fraction,
the $90\%$ confidence interval is $f_* = 4-11\%$. This value is slightly lower than other estimates that use the Salpeter IMF\cite{madaudickinson2014} because the noisy data points at $z<0.1$ happen to be above the global fit. A Chabrier mass function for the same data produces an integrated mass that is $\sim$\,1.7 times smaller, i.e., $f_* = 2.5-6.5\%$ at $90\%$ confidence. On top of the IMF uncertainty, 
a discrepancy exists between the value of $\rho_*$ from 
modelling mass to light ratios vs. integrating the cosmic 
star formation rate\cite{madaudickinson2014}, the latter being about 
$40\%$ higher. However, it may now be resolved\cite{Leja2022}.

We have used our measurements of the 
cosmic baryon budget to bound the total stellar 
mass in the Universe, and therefore constrain 
the mean IMF. Noting that $f_* = 1 - f_{d} - f_{cold}$, we place an upper 
limit on the stellar baryon fraction at low redshifts 
by deriving a lower limit on $f_d + f_{cold}$, which includes our FRB results, the ISM, and the cold CGM. We find $f_*\leq9\%$ and therefore $\rho_* \leq 5.6\times10^{8}\,M_\odot\,\mathrm{Mpc}^{-3}$ with
$90\%$ confidence.
We then tether our $\rho_*$ upper-limit to a recent estimate of the stellar mass that used a Chabrier IMF\cite{Leja2020}. We are able to rule out the Salpeter IMF with a low-mass cut-off below $0.10\,M_\odot$.

\clearpage

\noindent {\bf Data Availability Statement.} The FRB data 
presented here is publicly available in a CSV file at the following link:

\noindent https://github.com/liamconnor/frb\_baryon\_connor2024/blob/main/data/frbsample\_connor0924.csv

\noindent {\bf Code Availability Statement.} We have created 
a reproduction package for our work that includes all 
code used for our data analysis and the production of 
each figure. We have placed this code on GitHub at https://github.com/liamconnor/frb\_baryon\_connor2024.

\noindent {\bf Acknowledgements} We thank Fabian Walter, Xavier Prochaska, and Martijn Oei for informative conversations. We also
thank Dylan Nelson and Charles Walker for their considerable help with IllustrisTNG.

\noindent {\bf Author contributions.}
V.R. and G.Ha. led the development of the DSA-110. 
D.H., M.H., J.L., P.R., S.W., and D.W. contributed 
to the construction of the DSA-110.
L.C. conceived of and performed the analysis techniques 
for studying the FRB sample, as well as the multiwavelength baryon analysis. 
L.C. led the writing of the manuscript, 
with assistance from all coauthors. 
K.S., V.R., L.C., C.L., J.S., J.F., N.K., and M.S. all 
conducted the optical/IR follow-up observations presented 
in this work. K.S. and V.R. undertook the majority of the 
optical/IR host galaxy data analysis and interpretation.
V.R., C.L., L.C., G.He., and R.H. developed the 
software pipeline for detecting FRBs on the DSA-110. R.K. 
led the investigation of ray-tracing in the IllustrisTNG
simulation.

\noindent {\bf Competing interests statement.} The authors declare that they have no competing interests, financial or otherwise.

\noindent {\bf Correspondence.} Correspondence and request for materials should be addressed to L. Connor (email: liam.connor@cfa.harvard.edu)

\clearpage

\begin{table*}[ht!]
\centering
\begin{tabular}{lrrrrl}
\toprule
\textbf{Name} & $\rm{\mathbf{DM_{obs}}}$ &  $\rm{\mathbf{DM_{ex}}}$  & \textbf{Redshift} & $\rm{\mathbf{DM_{MW}}}$  & \textbf{Survey} \\
\midrule
FRB 20220204A & 612.20 & 561.50 & 0.4000 & 50.7 & DSA-110 \\
FRB 20220207C & 262.30 & 186.30 & 0.0430 & 76.0 & DSA-110 \\
FRB 20220208A & 437.00 & 335.40 & 0.3510 & 101.6 & DSA-110 \\
FRB 20220307B & 499.15 & 371.00 & 0.2507 & 128.2 & DSA-110 \\
FRB 20220310F & 462.15 & 415.90 & 0.4790 & 46.3 & DSA-110 \\
FRB 20220319D & 110.95 & -28.80 & 0.0111 & 139.8 & DSA-110 \\
FRB 20220330D & 468.10 & 429.50 & 0.3714 & 38.6 & DSA-110 \\
FRB 20220418A & 623.45 & 586.80 & 0.6220 & 36.7 & DSA-110 \\
FRB 20220506D & 396.93 & 312.40 & 0.3005 & 84.5 & DSA-110 \\
FRB 20220509G & 269.50 & 213.90 & 0.0894 & 55.6 & DSA-110 \\
FRB 20220726A & 686.55 & 597.00 & 0.3610 & 89.5 & DSA-110 \\
FRB 20220825A & 651.20 & 572.70 & 0.2414 & 78.5 & DSA-110 \\
FRB 20220831A & 1146.25 & 1019.50 & 0.2620 & 126.7 & DSA-110 \\
FRB 20220914A & 631.05 & 576.40 & 0.1138 & 54.7 & DSA-110 \\
FRB 20220920A & 315.00 & 275.10 & 0.1585 & 39.9 & DSA-110 \\
FRB 20221012A & 442.20 & 387.80 & 0.2840 & 54.4 & DSA-110 \\
FRB 20221027A & 452.50 & 405.30 & 0.2290 & 47.2 & DSA-110 \\
FRB 20221029A & 1391.05 & 1347.10 & 0.9750 & 43.9 & DSA-110 \\
FRB 20221101B & 490.70 & 359.50 & 0.2395 & 131.2 & DSA-110 \\
FRB 20221113A & 411.40 & 319.70 & 0.2505 & 91.7 & DSA-110 \\
FRB 20221116A & 640.60 & 508.30 & 0.2764 & 132.3 & DSA-110 \\
% FRB 20221203A & 602.25 & 518.80 & 0.5100 & 83.4 & DSA-110 \\
FRB 20221219A & 706.70 & 662.30 & 0.5540 & 44.4 & DSA-110 \\
FRB 20230124A & 590.60 & 552.10 & 0.0940 & 38.5 & DSA-110 \\
FRB 20230216A & 828.00 & 789.50 & 0.5310 & 38.5 & DSA-110 \\
FRB 20230307A & 608.90 & 571.30 & 0.2710 & 37.6 & DSA-110 \\
FRB 20230501A & 532.50 & 406.90 & 0.3010 & 125.6 & DSA-110 \\
FRB 20230521B & 1342.90 & 1204.10 & 1.3540 & 138.8 & DSA-110 \\
FRB 20230626A & 451.20 & 412.00 & 0.3270 & 39.2 & DSA-110 \\
FRB 20230628A & 345.15 & 306.00 & 0.1265 & 39.1 & DSA-110 \\
FRB 20230712A & 586.96 & 547.80 & 0.4525 & 39.2 & DSA-110 \\
FRB 20230814B & 696.40 & 591.50 & 0.5535 & 104.9 & DSA-110 \\
FRB 20231120A & 438.90 & 395.10 & 0.0700 & 43.8 & DSA-110 \\
FRB 20231123B & 396.70 & 356.50 & 0.2625 & 40.2 & DSA-110 \\
FRB 20231220A & 491.20 & 441.30 & 0.3355 & 49.9 & DSA-110 \\
FRB 20240119A & 483.10 & 445.20 & 0.3700$^\dagger$ & 37.9 & DSA-110 \\
FRB 20240123A & 1462.00 & 1371.70 & 0.9680 & 90.3 & DSA-110 \\
FRB 20240213A & 357.40 & 317.30 & 0.1185 & 40.1 & DSA-110 \\
FRB 20240215A & 549.50 & 501.50 & 0.2100 & 48.0 & DSA-110 \\
FRB 20240229A & 491.15 & 453.20 & 0.2870$^\dagger$ & 37.9 & DSA-110 \\
\bottomrule
\end{tabular}
\end{table*}

\clearpage

\begin{table*}[ht!]
\centering
\begin{tabular}{lrrrrl}
\toprule
\textbf{Name} & $\rm{\mathbf{DM_{obs}}}$ &  $\rm{\mathbf{DM_{ex}}}$  & \textbf{Redshift} & $\rm{\mathbf{DM_{MW}}}$  & \textbf{Survey} \\
\midrule
FRB20121102A & 558.10 & 369.70 & 0.1927 & 188.4 & Arecibo \\
FRB20171020A & 114.10 & 77.60 & 0.0087 & 36.5 & ASKAP \\
FRB20180301A & 536.00 & 384.40 & 0.3304 & 151.6 & Parkes \\
FRB20180916B & 348.76 & 149.90 & 0.0337 & 198.9 & CHIME \\
FRB20180924A & 362.16 & 321.70 & 0.3212 & 40.4 & CHIME \\
FRB20181030A & 103.50 & 62.40 & 0.0039 & 41.1 & CHIME \\
FRB20181112A & 589.27 & 547.50 & 0.4755 & 41.7 & ASKAP \\
FRB20190102C & 364.55 & 307.10 & 0.2912 & 57.4 & ASKAP \\
FRB20190520B & 1210.30 & 1150.20 & 0.2410 & 60.1 & FAST \\
FRB20190523A & 760.80 & 723.60 & 0.6600 & 37.2 & DSA-10 \\
FRB20190608B & 340.05 & 302.90 & 0.1178 & 37.2 & ASKAP \\
FRB20190711A & 587.90 & 531.50 & 0.5217 & 56.4 & ASKAP \\
FRB20190714A & 504.13 & 465.70 & 0.2365 & 38.4 & ASKAP \\
FRB20191001A & 507.90 & 463.70 & 0.2340 & 44.2 & ASKAP \\
FRB20191228A & 298.00 & 265.10 & 0.2430 & 32.9 & ASKAP \\
FRB20200120E & 87.82 & 47.10 & 0.0008 & 40.8 & CHIME \\
FRB20200430A & 380.00 & 352.90 & 0.1610 & 27.1 & ASKAP \\
FRB20200906A & 577.84 & 542.00 & 0.3688 & 35.8 & ASKAP \\
FRB20201123A & 433.55 & 181.10 & 0.0507 & 252.5 & MeerKAT \\
FRB20201124A & 411.00 & 271.10 & 0.0982 & 139.9 & CHIME \\
FRB20210117A & 728.95 & 694.70 & 0.2145 & 34.3 & ASKAP \\
FRB20210320C & 384.59 & 345.40 & 0.2797 & 39.2 & ASKAP \\
FRB20210410D & 575.00 & 518.80 & 0.1415 & 56.2 & MeerKAT \\
FRB20210807D & 251.30 & 130.00 & 0.1293 & 121.3 & ASKAP \\
FRB20211127I & 234.97 & 192.50 & 0.0469 & 42.5 & ASKAP \\
FRB20211203C & 635.00 & 571.40 & 0.3439 & 63.6 & ASKAP \\
FRB20211212A & 209.00 & 170.20 & 0.0707 & 38.8 & ASKAP \\
FRB20220105A & 580.00 & 558.10 & 0.2785 & 21.9 & ASKAP \\
FRB20220610A & 1458.10 & 1427.20 & 1.0150 & 30.9 & ASKAP \\
\bottomrule
\end{tabular}
\caption{The basic properties of cosmological FRBs with host galaxy redshifts. All DM values have units pc\,cm$^{-3}$. $\dagger$ indicates that the redshift 
is photometric. All others are spectroscopic.\label{table:frbs}}
\label{Table}
\end{table*}

\begin{table*}[ht!]
    \setlength{\tabcolsep}{3.5pt}
    \centering
    \captionsetup{labelformat=tablelabel}
    \caption{Basic FRB and host properties of 9 new sources.}
    \footnotesize
    \begin{tabular}{lrrrrrlrrr}
        \toprule
        
        FRB & RA (FRB) & Decl. (FRB) & RA (Host) & Decl. (Host) & P$_\mathrm{host}$ & $z$ & M$_{\mathrm{AB}}$ & Filter & E(B-V) \\
        & [J2000] & [J2000] & [J2000] & [J2000] & & & [mag] & & \\

        \hline

        FRB\,\frbannie & 11:04:40.39 & +74:04:31.40 & 11:04:40.27 & +74:04:28.86 & 0.99 & 0.1185 & 19.33 & BASS $r$ & 0.074 \\

        (Annie) & $\pm$ 1.00 & $\pm$ 0.60 & & & & & $\pm$ 0.04 & & \\

        FRB\,\frbbubble & 17:53:45.90 & +70:13:56.50 & 17:53:45.97 & +70:13:56.18 & 0.99 & 0.2100 & 20.36 & BASS $r$ & 0.037 \\

        (Bubble) & $\pm$ 0.80 & $\pm$ 0.50 & & & & & $\pm$ 0.02 & & \\

        FRB\,\frbada & 22:34:46.93 & +70:32:18.40 & 22:34:47.13 & +70:32:17.40 & 0.99 & 0.2620 & 22.70 & LRIS $R$ & 0.598 \\

        (Ada) & $\pm$ 1.03 & $\pm$ 0.67 & & & & & $\pm$ 0.06 & & \\

        FRB\,\frbgemechu & 08:15:38.09 & +73:39:35.70 & 08:15:38.47 & +73:39:34.76 & 0.99 & 0.3355 & 20.49 & PS1 $r$ & 0.806 \\

        (Gemechu) & $\pm$ 0.70 & $\pm$ 0.50 & & & & & $\pm$ 0.04 & & \\

        FRB\,\frbcasey & 11:19:56.05 & +70:40:34.40 & 11:19:56.48 & +70:40:34.67 & 0.99 & 0.2870 & 21.12 & BASS $r$ & 0.018 \\

        (Casey) & $\pm$ 0.80 & $\pm$ 0.60 & & & & & $\pm$ 0.06 & & \\

        FRB\,\frbnikhil & 14:57:52.12 & +71:36:42.33 & 14:57:53.01 & +71:36:40.99 & 0.98 & 0.3760 & 21.20 & BASS $r$ & 0.024 \\

        (Nikhil) & $\pm$ 1.30 & $\pm$ 0.70 & & & & & $\pm$ 0.02 & & \\

        % FRB\,\frbwilhelm & 21:00:31.09 & +72:02:15.22 & 21:00:30.83 & +72:02:15.65 & 1.00 & 0.5100 & 22.32 & LRIS $r$ & 0.773 \\

        % (Wilhelm) & $\pm$ 0.46 & $\pm$ 0.56 & & & & & $\pm$ 0.04 & & \\

        FRB\,\frbjohndoe & 22:23:53.94 & +73:01:33.26 & 22:23:54.26 & +73:01:32.77 & 0.99 & 0.5530 & 22.90 & DEIMOS $R$ & 0.650 \\

        (Johndoe) & $\pm$ 1.70 & $\pm$ 0.39 & & & & & $\pm$ 0.01 & & \\

        FRB\,\frbpushkin & 04:33:03.00 & +71:56:43.02 & 04:33:03.01 & +71:56:43.20 & 0.99 & 0.9680 & 21.91 & PS1 $r$ & 0.806 \\

        (Pushkin) & $\pm$ 1.30 & $\pm$ 0.80 & & & & & $\pm$ 0.04 & & \\

        FRB\,\frbbruce & 23:24:08.64 & +71:08:16.91 & 23:24:07.88 & +71:08:17.59 & 0.95 & 1.3540 & 21.15 & WIRC $J$ & 0.970 \\

        (Bruce) & $\pm$ 1.20 & $\pm$ 0.60 & & & & & $\pm$ 0.12 & & \\

        \hline
    \end{tabular}
    \label{table:basic_frb_properties}
\end{table*}

\clearpage

\end{document}